\documentclass[amsmath,amssymb,aps,prb,twocolumn,floatfix,showpacs,superscriptaddress]{revtex4-1}

\usepackage{graphicx}
\usepackage{dcolumn}
\usepackage{bm}
\usepackage{subfigure}
\usepackage{epstopdf}
\usepackage{enumerate}

\newcommand{\be}{\begin{equation}}
\newcommand{\ee}{\end{equation}}
\newcommand{\ba}{\begin{array}}
\newcommand{\ea}{\end{array}}
\newcommand{\bea}{\begin{eqnarray}}
\newcommand{\eea}{\end{eqnarray}}

\def\temp(#1){\langle #1\rangle}

\def\tempp(#1){\langle {#1}|}

\def\temppp(#1){|#1\rangle}

\def\ttt(#1,#2){\left(\!\!\ba{c} {#1}\\{#2}\ea\!\!\right)}
\def\spin#1{\ttt#1}

\def\tttt(#1,#2){\left(\!\!\ba{cc} {#1} & {#2}\ea\!\!\right)}
\def\hspin#1{\tttt#1}

\pdfpageattr {/Group << /S /Transparency /I true /CS /DeviceRGB>>}

\begin{document}

\title{Nonlinear response of a ballistic graphene transistor with an ac-driven gate:\\ high harmonic generation and THz detection}

\author{Y. Korniyenko}
\affiliation{Department of Microtechnology and Nanoscience - MC2,
Chalmers University of Technology, SE-412 96 G\"oteborg, Sweden}

\author{O. Shevtsov}
\affiliation{Department of Physics \& Astronomy, Northwestern University,
Evanston, IL 60208}

\author{T. L\"ofwander}
\affiliation{Department of Microtechnology and Nanoscience - MC2,
Chalmers University of Technology, SE-412 96 G\"oteborg, Sweden}

\date{\today}

\begin{abstract}

We present results for time-dependent electron transport in a ballistic graphene field-effect transistor with an ac-driven gate.
Nonlinear response to the ac drive is derived utilizing Floquet theory for scattering states
in combination with Landauer-B\"uttiker theory for transport.
We identify two regimes that can be useful for applications:
(i) low and (ii) high doping of graphene under source and drain contacts, relative to the doping level in the graphene channel,
which in an experiment can be varied by a back gate.
In both regimes, inelastic scattering induced by the ac drive can excite quasi-bound states in the channel
that leads to resonance promotion of higher order sidebands.
Already for weak to intermediate ac drive strength, this leads to a substantial change in the direct current between source and drain.
For strong ac drive with frequency $\Omega$, we compute the higher harmonics of frequencies $n\Omega$ ($n$ integer)
in the source-drain conductance.
In regime (ii), we show that particular harmonics (for instance $n=6$) can be selectively enhanced
by tuning the doping level in the channel or by tuning the drive strength.
We propose that the device operated in the weak-drive regime can be used to detect THz radiation,
while in the strong-drive regime it can be used as a frequency multiplier.

\end{abstract}


\maketitle

\section{Introduction}\label{sec:intro}

Graphene for analogue high-frequency electronics has been the focus of intense research the last few years,
and is one of the focus areas in the recently published graphene roadmap\cite{AndreaCFerrari:2015co}.
Two-dimensionality of the material, high carrier mobility, gate-tunable charge density,
and a unique band structure with massless Dirac electrons are a few of the properties that make graphene
a promising material in this context\cite{Schwierz:2010ix,Palacios:2010dw,Glazov:2014cp,Otsuji:2012hn,Koppens:2014dy}.
Examples of devices already produced, with competitive figures of merits, are
field-effect transistors\cite{2012PNAS..10911588C},
frequency doublers\cite{Wang:2009jq},
frequency mixers\cite{Habibpour:fz},
and detectors\cite{Vicarelli:2012ch,Mittendorff:2013gk,Cai:2014hh,Zak:2014gc}.

The electronic mobility has been constantly improving and ballistic electron transport is today studied intensively.
Ballistic transport allows for development of massless Dirac electron optics, which is the graphene analogue of usual optics.
Electron optics effects that have been observed include Fabry-P\'erot interferences and snake states\cite{Rickhaus:2015cp},
Veselago lensing\cite{Anonymous:N5gujY47},
and so-called whispering gallery modes in circular $p$-$n$ junctions\cite{2015Sci...348..672Z}.

For ballistic devices, evidence of hydrodynamic behavior has been recently presented:
viscous electron backflow\cite{Bandurin:2016cp} and breakdown of the Wideman-Franz law\cite{Crossno:2016iy,Ghahari:2016df}.
This indicates that due to the long elastic mean free path, and slow electron-phonon relaxation below room temperature,
electron-electron interactions can be the most dominant scattering channel within a certain temperature window.
However, at sufficiently low temperatures (below 100 K) electron-electron interactions also become weak and, ultimately, at lower temperature,
transport is truly ballistic over long ($\mu m$) length scales.

Improved mobility (possibly reaching ballistic transport) is a necessary condition for the development of high-frequency devices.
There has therefore been a broad interest in the theory of time-dependent transport in graphene in the ballistic transport regime,
including quantum pumping\cite{Prada:2009jo,FoaTorres:2011kf,SanJose:2011fe,SanJose:2012bw},
nonlinear electromagnetic response\cite{Mikhailov:2007ft,Mikhailov:2008ig,Syzranov:2008dg,Calvo:2012bg,AlNaib:2014jj,Sinha:2012fx},
and photon-assisted tunneling\cite{Trauzettel:2007kq,Zeb:2008ka,Rocha:2010ho,Savelev:2012dg,Lu:2012it,Szabo:2013bn,Zhu:2015bd}.
In the non-classical regime, when the energy scale $\hbar\Omega$, set by the drive frequency $\Omega$
($\hbar$ is Planck's constant divided by $2\pi$),
and the Fermi energy $E_F$, measured relative to the charge-neutrality point, are of comparable magnitude,
a variety of interesting quantum mechanical interference and resonance effects become important.
In a recent paper\cite{Korniyenko:2016ct} we have studied in detail
a Fano resonance\cite{BAGWELL:1992wx,Lu:2012it,Szabo:2013bn,Zhu:2015bd}
induced by a quasibound bound state on the top gate barrier.
We showed how it could be utilized to develop a frequency doubler for weak or moderate ac drive strength.
In this paper we extend this study to include a more realistic doping profile across the device as well as strong ac drive.
Within a fully quantum mechanical treatment based on Floquet theory  and Landauer-B\"uttiker scattering
theory\cite{BAGWELL:1992wx,1998PhRvB..5812993P,Platero:2004ep,2005PhR...406..379K},
we show how Fano resonances as well as resonant tunneling can be utilized for detection of high-frequency radiation in the THz range
or to generate high harmonics of the ac signal.

The outline of the paper is as follows.
In Section~\ref{sec:model} we give details of the model and the methods of calculations.
This section also includes a characterization of the dc regime as a prologue to the discussions of time-dependent transport in the following chapter,
as well as a detailed discussion of the relation between the different parameters of the model and various possible transport regimes.
In Section~\ref{sec:weak_drive} we present results for the weak ac drive regime, with focus on high-frequency radiation detection.
In Section-\ref{sec:strong_drive} we present result for the strong ac drive regime, with focus on high harmonic generation.
Section~\ref{sec:summary} summarizes the paper.
A few technical results are collected in the Appendix.

\section{Model}\label{sec:model}

\begin{figure}[t]
\includegraphics[width=0.9\columnwidth]{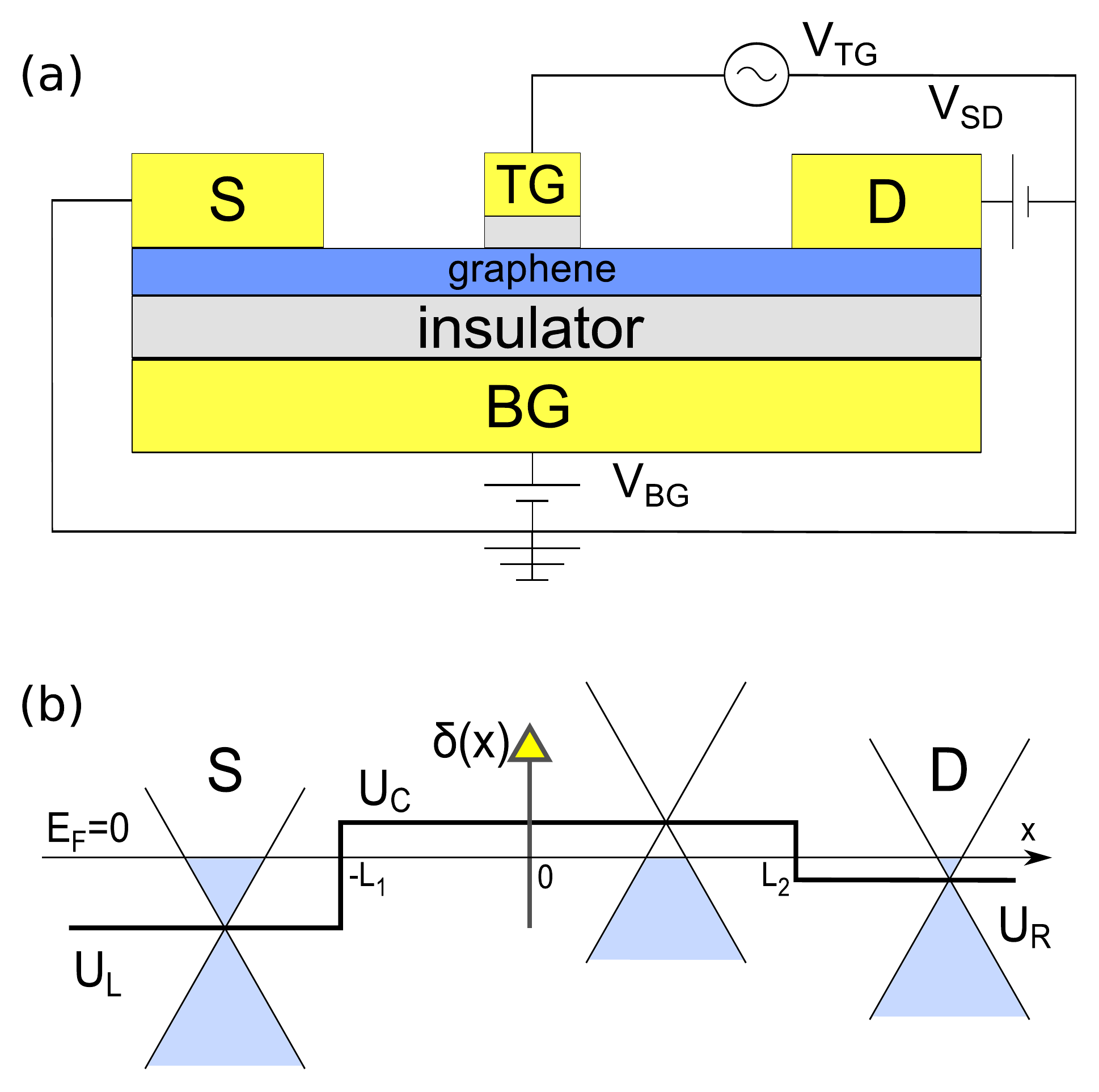}
\caption{(a) Schematics of a graphene field effect transistor, where a back gate (BG) controls doping of the channel,
a small source (S) - drain (D) bias is applied to generate the current, which is controlled by the top gate (TG) dc and ac signals.
(b) Potential landscape, including doping of the leads by the source and drain metallic electrodes.}
\label{tr}
\end{figure}

Our goal is to establish a relation between intrinsic electronic transport properties of a ballistic graphene transistor,
depicted in Fig.~\ref{tr}(a), and experimentally controllable physical parameters.
Extrinsic (parasitic) effects due to eventual surrounding circuit elements, must be dealt with when doing experiments,
but can be neglected in an attempt to describe the intrinsic properties.
We make a minimal model based on a number of assumptions that we outline in the following.

First, we assume that the contacts and gates are ideal,
such that they can be described by the potential landscape sketched in Fig.~\ref{tr}(b).
We take into account that the source and drain metallic contacts dope graphene underneath due to work function mismatches.
The doping levels, set by $U_L$ and $U_R$, in the graphene source and drain areas are thereby pinned \cite{Huard:2008hx}.
On the other hand, in the transistor channel region, $x\in[-L_1,L_2]$,
the doping level can be tuned by the back gate potential.
We define the channel Dirac point energy by setting $E_D=U_C$ (assuming absence of e-h puddles),
where $U_C$ can be tuned by the  back gate.
Since we measure energies with respect to the Fermi level $E_F=0$ (aligned with the metallic contact Fermi energies),
the Dirac point in the channel region is aligned with the Fermi energy for $U_C=0$ (the channel is then charge neutral).
In summary, the doping profile sketched in Fig.~\ref{tr}(b) is given by
\bea
U(x) &=& U_L\theta(-L_1-x) + U_R\theta(x-L_2) \nonumber\\
&&+\,U_C\left[ \theta(x+L_1)-\theta(x-L_2)\right].
\label{doping_profile}
\eea

We assume that the top gate is wide on the scale of the C-C bond length $a_{cc}$,
but short on the scale that the envelope of the Dirac electron wavefunction varies,
which is given by $\lambda_D=\hbar v_F/(E-U_C)$, where $v_F$ is the Fermi velocity.
For energies $E$ near the Dirac point in the channel, we have $\lambda_D\gg a_{cc}$.
Based on the same arguments we assume that the doping level is changing slowly near the contacts on the $a_{cc}$ scale
but fast on the scale of $\lambda_D$.
These assumptions mean that we can neglect intervalley scattering in the problem and consider only one valley.
For transport quantities, a factor two for valley degeneracy is included in addition to the factor two for spin degeneracy.
The above assumptions also allow us to use step functions for the doping profile, as in Eq.~(\ref{doping_profile}),
and a delta barrier model for the top gate potential.
The effective low-energy Hamiltonian then has the form
\be
\mathcal{H}=-i\sigma_x\nabla_x+\sigma_y k_y+\left[Z_0+Z_1\cos(\Omega t)\right]\delta(x)+U(x),
\label{Hamiltonian}
\ee
where we have set the Fermi velocity in graphene equal to unity, $v_F=1$, and $\hbar=1$.
The Pauli matrices in pseudo-spin space (A-B sublattices) are denoted by $\sigma_x$ and $\sigma_y$.
We assume the device to be very wide and translationally invariant along $y$.
Thus any edge effects are negligibly small and transverse momentum $k_y$ is (approximately) conserved.
Above, $Z_{0}$ and $Z_{1}$ are respectively static and dynamic parts of the drive applied at the top gate.
The delta function description of the top gate barrier is obtained as a limiting case of a very high $V\rightarrow\infty$
and narrow $D\rightarrow 0$ square barrier, with the product (barrier strength) $V D=Z$ constant.
Note that this theory for the Dirac quasiparticle envelope wavefunction holds as long as $a_{cc}\ll D\ll\lambda_D$.

Wave function solutions have to satisfy the time-dependent Dirac equation
\begin{equation}
\mathcal{H}\psi(x,k_y,t)=i\partial_t\psi(x,k_y,t).
\label{Dirac_eq}
\end{equation}
The harmonic potential, with frequency $\Omega$, in the Hamiltonian $\mathcal{H}$ in Eq.~(\ref{Hamiltonian})
allows us to use a Fourier decomposition and construct a Floquet ansatz,
\begin{equation}
\psi(x,k_y,t) = e^{-iEt} \sum\limits_{n=-\infty}^{\infty} \psi_n(x,k_y,E) e^{-in\Omega t},
\end{equation}
where amplitudes at sideband energies $E_n=E+n\Omega$ ($n$ integer) are the result of the charge carrier picking up (or giving up)
energy quanta $n\Omega$ from the oscillating barrier.
The quasi-energy $E$ is set by the energy of the particle incident from the source electrode in the scattering problem.
When plugged into Eq.~(\ref{Dirac_eq}) it yields a set of coupled differential equations for sideband amplitudes $\psi_n(x,k_y,E)$.
The solutions can be derived in a straightforward manner by wavefunction matching and collected into a Floquet scattering matrix
describing scattering of a quasiparticle incoming from left or right reservoir at energy $E$ and transverse momentum $k_y$.
We have collected all the key steps of the derivation in the Appendix.
The reflection amplitudes $r_n(k_y,E)$ are given in Eq.~(\ref{r_n}) and the transmission amplitudes $t_n(k_y,E)$
are given in Eq.~(\ref{t_n}).

Following the Landauer-B\"uttiker scattering approach, the Floquet scattering matrix can be used to compute
the time-dependent conductance $G(t)$ between source and drain.
The conductance is computed in linear response to the source-drain voltage $V_{SD}$,
but in non-linear response to the oscillating top gate potential, described by its drive strength $Z_1$ and frequency $\Omega$.
The conductance is also a function of the static potential landscape, described by $U(x)$,
as well as the static top gate potential quantified by its barrier strength $Z_0$.
We derived the general formula for $G(t)$ in Ref.~\onlinecite{Korniyenko:2016ct}.
Here we choose to present results for the linear conductance in the right lead at $x=L_2^+$, i.e. at the interface with the channel region.
The expression for the conductance (per unit length in the transverse direction) at zero temperature is
then\footnote{We note that for a system translationally invariant in the transverse direction, one has to compute conductance per unit length. In our previous work, c.f. Eqs.~(D10)-(E2) in Ref.~\onlinecite{Korniyenko:2016ct}, we missed the $1/2\pi$ prefactor associated with $k_y$-integration in the current and conductance formulas, which we include here [see Eq. (\ref{eq:conductance})].
The main results of Ref.~\onlinecite{Korniyenko:2016ct} are not affected, although the scales in Fig.~3(b) and 4 should include this prefactor.}
\begin{widetext}
\begin{eqnarray}
G(E_F,t) &=& \sum_{n=-\infty}^{\infty} G_{n}(E_F)e^{-in\Omega t},\\
G_{n}(E_F) &=& G_{-n}^*({E_F}),\\
G_n(E_F) &=& \frac{4e^2}{h} \int_{-\infty}^{\infty} \frac{dk_y}{2\pi}
\sum_{m=-\infty}^{\infty} \frac{\eta^*(k_y,E_m-U_R)+\eta(k_y,E_{n+m}-U_R)}{2\sqrt{v(k_y,E_m-U_R)v(k_y,E_{n+m}-U_R)}}\nonumber\\
&&\times\,\left.e^{i\left[\kappa(k_y,E_m-U_R)-\kappa(k_y,E_{n+m}-U_R)\right]L_2}\,
t_m^{\dagger}(k_y,E) t_{n+m}(k_y,E) \right |_{E=E_F},\label{eq:conductance}
\end{eqnarray}
\end{widetext}
where $\eta(k_y,E)$, $v(k_y,E)$, and $\kappa(k_y,E)$ are defined in Eq.~(\ref{velocity}).
The factor with velocities appears here because we utilize a scattering basis
where elementary waves in the leads carry unit probability flux. This guarantees that the scattering matrix
coupling incoming and outgoing waves in the leads is unitary.
The phase of the conductance components $G_n$ for $n \neq 0$ is unimportant for our discussion and we will present results for $|G_n|$ below.
Note that in the static case (i.e. $Z_1=0$), the factor with velocities as well as the phase factor both reduce to unity
and the usual Landauer-B\"uttiker formula for dc conductance simply in terms of transmission is obtained.

\begin{figure}[t]
\includegraphics[width=\columnwidth]{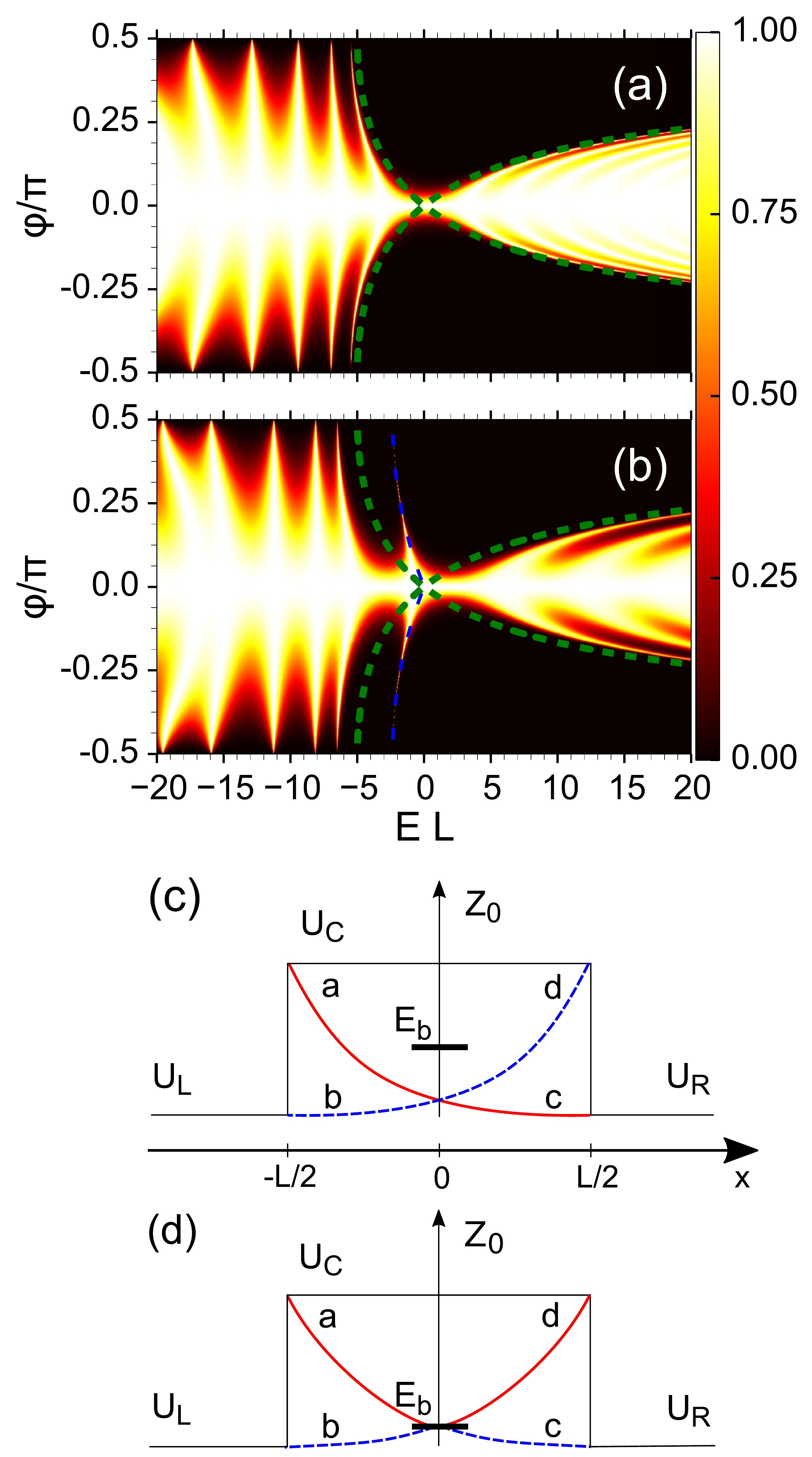}
\caption{(a) Dc transmission probability as a function of energy and incidence angle.
Electrodes are doped at $U=-10/L$, while the gate potentials are zero such that $Z_0=0$ and $U_C=0$.
Green dashed lines indicate boundaries to evanescent regions $|\varphi|>\varphi_{c}$.
(b) Transmission probability in the presence of a top gate dc potential, $Z_0=0.4\pi$.
Blue long-dashed lines indicate resonant tunneling. (c) and (d) display the connection between evanescent waves
at the delta barrier for off- and on-resonance tunneling, respectively.}
\label{fig:contacts}
\end{figure}

In the rest of the paper we shall report results for a symmetric setup with $L_1=L_2=L/2$ and symmetric doping profile $U_L=U_R=U$.
The transmission probabilities are computed for zero back gate voltage, i.e. $U_C=0$, as a function of energy $E$
and transverse momentum $k_y=|E-U|\sin\varphi$ parametrized by an impact angle $\varphi$.
This means that $E=0$ corresponds to transmission at the Dirac point in the channel region.
This is a conventional way to present transmission through a potential landscape.
On the other hand, the zero-temperature linear conductance, computed via Eq.~(\ref{eq:conductance}),
shall be presented as a function of the channel doping $U_C$ (the position of the Dirac point energy $E_D$).
In an experiment, the channel doping level can be tuned by the back gate voltage $V_{BG}$.
Since the Fermi energy is pinned to the metallic source and drain contact Fermi energies,
the radius of the Dirac cone in the graphene leads is constant, set by the doping level $U$,
while the radius in the channel is given by $U_C$ and varies with back gate voltage.
This choice should correspond to the experimental situation.

\subsection{dc characteristics}\label{sec:dc}

We start by analyzing the static case ($Z_1=0$) in order to set the stage for the signatures of the time-dependent drive
that we will study in the following sections.
It is useful to first look at the case with no applied top gate potential $Z_0=0$, thereby highlighting the effect of the inhomogeneous
doping profile. In fact, $U(x)$ describes a square barrier across the channel of width $L=L_1+L_2$.
We plot the transmission probability $T_0(E,\varphi)$ in Fig.~\ref{fig:contacts}(a).
The transmission amplitude is governed by pseudospin matching between regions with different doping.
For small angles $\varphi$ the mismatch is negligibly small, thus transmission approaches unity (Klein tunneling).
The peaks in transmission for large angles $\varphi$ and negative energies $E<-5/L$ in Fig.~\ref{fig:contacts}(a)
are analogous to Fabry-P\'erot fringes,
i.e. the result of wave interference between two partially reflecting mirrors (boundaries at the source and drain in this case).
A typical fringe oscillation period is of the order of $2\pi\hbar v_f/L$ (reinstating the units).
In addition to the two effects described above, there is a large region where transmission is largely suppressed.
It occurs when the waves in the channel region are evanescent.
Their longitudinal momentum component $\kappa(k_y,E)=\pm\sqrt{E^2-k_y^2}$ turns imaginary,
giving us a condition on the critical angle of incidence $\varphi_c$,
\be
\varphi_c = \arcsin\left| \frac{E}{E-U} \right|.
\label{eq:phi_c}
\ee
For any $|\varphi|>\varphi_c$ the waves injected from the electrodes are evanescent in the channel ($x\in [-L_1,L_2]$).
Note that Eq.~(\ref{eq:phi_c}) holds for $|E|<|E-U|$.
Otherwise there are no evanescent waves involved in transport and we may put $\varphi_c=\pi/2$.
The boundary between propagating wave transport and evanescent wave transport
is indicated by a green dashed line in Fig.~\ref{fig:contacts}(a).
The evanescent wave factor $\exp(-\sqrt{k_y^2-E^2}L)$ lowers the transmission probability in general.
However, for energies close to the Dirac point and small $k_y$ (or small $L$), this factor is still quite large
and evanescent waves can reach between the two contacts thus giving rise to large transmission probability.
Transport at $E=0$ is achieved exclusively through evanescent waves. This is the so-called pseudo-diffusive transport regime.\cite{Tworzydio:2006hw}

When we introduce the static top gate delta barrier potential, $Z_0\ne 0$, additional features appear in the transmission.
First, the Fabry-P\'erot oscillations are shifted due to an additional phase shift at the delta barrier, see Fig.~\ref{fig:contacts}(b).
More importantly, the delta barrier can host one bound state at energy
\be
\label{eq:bound}
E_b = U_C - \mbox{sgn}(Z_0) |k_y| \cos Z_0,
\ee
that we studied for $U(x)=0$ in Ref.~\onlinecite{Korniyenko:2016ct}.
In that case, the bound state does not affect dc transport properties, but can be excited by ac drive.
Here, for finite electrode doping $U\ne 0$, the bound state can be excited already in dc.
In fact, in this case it is not a true bound state, rather a quasibound state with evanescent waves in the channel region
connected to propagating waves in the leads.
In Fig.~\ref{fig:contacts}(b) we see that the resonance in $T_0(E,\varphi)$ originates at $E=0$ 
and then disperses with the angle of incidence $\varphi$.
The resonance can be understood in analogy with widely studied resonant double barrier tunnelling\cite{1984PhRvB..29.1970R,1987ApPhA..42..245Y}
in Schr\"odinger quantum mechanics. In the analogy, the two barriers correspond in our case to the two channel regions between
the contacts and the top gate delta barrier, and the resonant level between the barriers corresponds in our case to the quasibound state
in the delta barrier.
A complimentary point of view of the resonance can be found in the equations, see Appendix~\ref{Appendix:resonant_tunneling}.
Off resonance, exponentially decaying waves with amplitudes $a$ and $c$ are connected, as sketched in Fig.~\ref{fig:contacts}(c).
This results in an exponentially small transmission amplitude.
On the other hand, when the quasibound state is hit, the exponentially decaying wave with amplitude $a$ on one side of the delta barrier is coupled only to
exponentially rising solution with amplitude $d$ on the other side, as sketched in Fig.~\ref{fig:contacts}(d).
The exponential functions thereby cancel in the expression for the transmission which leads to resonance behavior
[c.f. Eq.~(\ref{eq:cross_connection})].

\begin{figure}[t]
\includegraphics[width=\columnwidth]{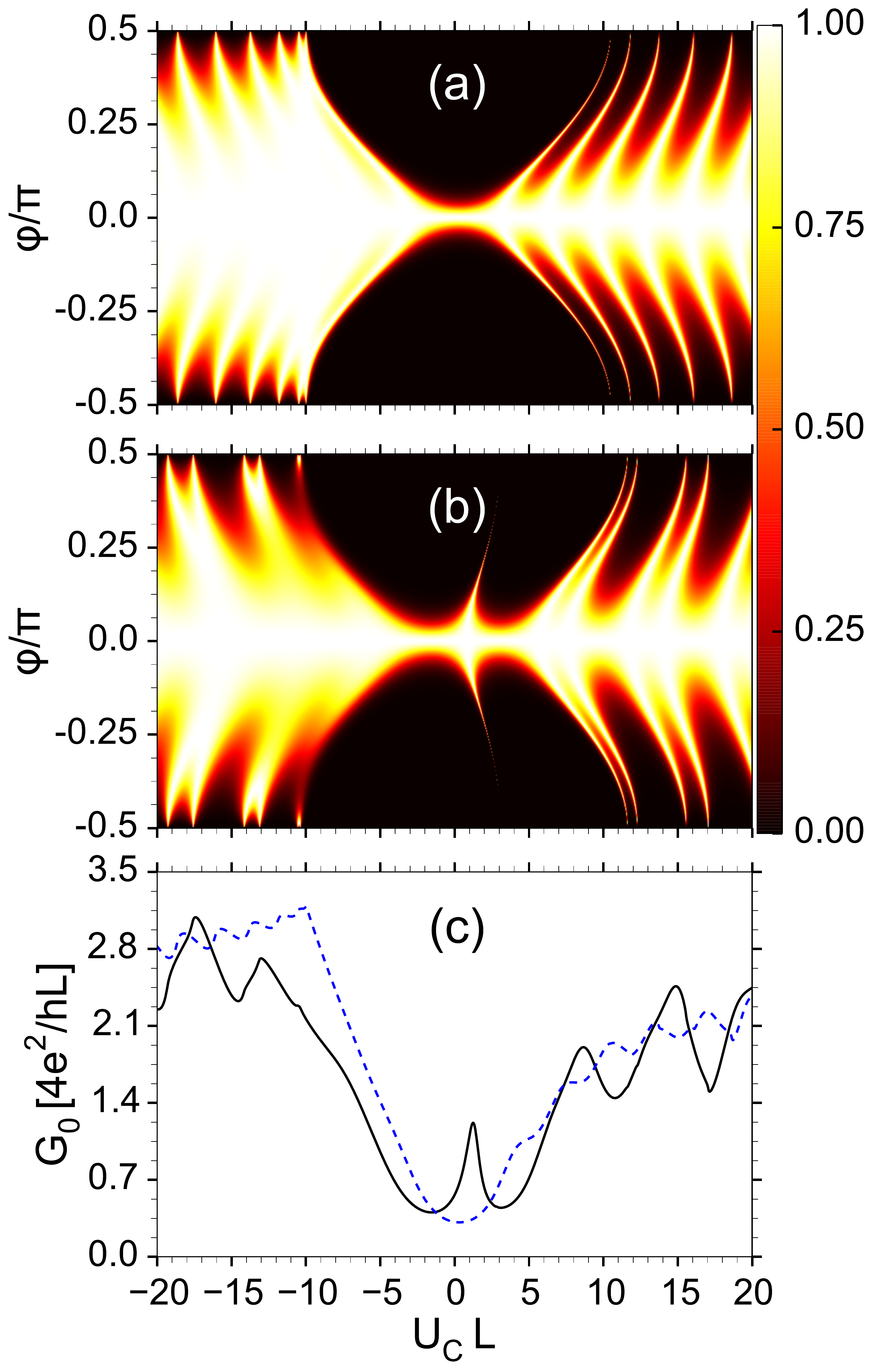}
\caption{(a) Transmission probability in the absence of top gate barrier ($Z_0=0$) for electrodes with pinned doping levels set by $U=-10/L$,
and varying channel doping level $U_C$, which defines the position of the Dirac point relative to the Fermi energy $E_F=0$.
(b) The transmission probability including a top gate barrier of strength $Z_0=0.4\pi$.
(c) Corresponding angle-integrated dc linear conductances for $Z_0=0$ (blue dashed line) and for $Z_0=0.4\pi$ (solid black line).}
\label{fig:G0}
\end{figure}

For the calculation of the conductance in Eq.~(\ref{eq:conductance}) we need to integrate the transmission probability over angles.
In an attempt to describe the typical experimental situation we assume that the Fermi energy in the device
and the doping levels in the leads are pinned by the Fermi energy in the metal contacts, while the back gate can be used
to tune the doping level in the channel.
The zero-temperature conductance as a function of $U_C$ is then computed by
integrating the transmission function $T(E,\varphi;U_L,U_R,U_C)$ over angles at fixed energy $E=0$ (Fermi energy) and fixed $U_L$ and $U_R$.
We plot the corresponding view of the angle-dependent transmission function in Fig.~\ref{fig:G0}(a)-(b).
Note that in Fig.~\ref{fig:contacts} we plotted $T(E,\varphi;U_L,U_R,U_C)$ as a function of $E$ and $\varphi$
for fixed $U_C=0$ and fixed $U_L$ and $U_R$.
The transmission function as viewed in Fig.~\ref{fig:G0}(a)-(b) corresponds to leads that are electron-doped (here $U=-10/L$).
Thus, both incoming waves and scattered waves in the leads are electron-like (n-type) at the Fermi energy $E_F=0$.
For $U_C<0$, we have electron-like waves at $E_F=0$ in the channel, while for $U_C>0$ we have hole-like waves in the channel (p-type).
Therefore, the Fabry-P\'erot interference patterns for positive $U_C$ (n-p-n junction) and negative $U_C$ (n-n'-n junction) are different.

In Fig.~\ref{fig:G0}(c) we present the dc conductance as a function of channel doping level $U_C$.
For $|U_C|<|U|$ we have mainly evanescent mode transport, while
for larger values of $|U_C|$ we find oscillations due to the Fabry-P\'erot interferences.
The resonance peak near $U_C=0$ (solid black line for finite $Z_0$) is due to the delta barrier induced quasibound state.

\subsection{Parameter regimes}\label{sec:parameters}

\begin{table}[t]
\caption{Energy scales within our model and relevant parameters they are determined by.}
\label{tab:scales}
\begin{tabular}{l|l}
\hline
$U$ & contact doping levels $U_R=U_L=U$\\
\hline
$U_C$ & channel doping level\\
\hline
$\hbar v_F/L$ & channel length $L=L_1+L_2$\\
\hline
$\hbar v_F/\Delta L$ & channel asymmetry $\Delta L=|L_1-L_2|$\\
\hline
$\hbar\Omega$ & drive frequency $\Omega$\\
\hline
\end{tabular}
\end{table}

Starting from the dc characterization above, we can identify several parameter regimes.
They can be described by different relations between the relevant energy scales in the problem, listed in Table~\ref{tab:scales}.
In the dc characterization above, we used $\hbar v_F/L$ as energy scale. Note that with $v_F=1=\hbar$,
energies are measured in units of $L^{-1}$.
In addition to the relations between the energy scales in Table~\ref{tab:scales},
we have to take into account the oscillating delta barrier strength $Z_1$.

The observed regimes in dc are (c.f. Fig.~\ref{fig:G0})
\begin{enumerate}[I.]
\item{$|U_C|\geq |U|$:} propagating wave transport
\begin{enumerate}
\item{$|U|\sim \hbar v_F/L$:} clearly visible Fabry-P\'erot interferences as a function of $U_C$ with period approximately given by $2\pi\hbar v_F/L$
\item{$|U|\gg \hbar v_F/L$:} very fast oscillations that in reality would be washed out by inhomogeneity or temperature smearing
\item{$|U|\ll \hbar v_F/L$:} the oscillations are too slow (on the scale of $U_C\sim U$) to be observed
\end{enumerate}
\item{$|U_C|\ll |U|$:} evanescent wave transport (pseudo-diffusive regime)
\begin{enumerate}
\item{$U\ll \hbar v_F/\Delta L$:} resonant tunneling is possible when the channel is not too asymmetric
\end{enumerate}
\end{enumerate}
The dc drive strength $Z_0$ sets the position of the quasibound state in resonant tunneling regime and shifts the Fabry-P\'erot oscillations,
but does not define a regime by itself.
We note that both the evanescent wave regime\cite{Danneau:2008kg} and
the Fabry-P\'erot regime\cite{Rickhaus:2015cp} have been observed experimentally.

Under ac drive we will in the next sections investigate the following regimes:
\begin{enumerate}[I.]
\setcounter{enumi}{2}
\item{$Z_1<1$:} Weak to intermediate drive
\begin{enumerate}
\item{$\hbar\Omega\gtrsim U$, low contact doping; with I.a above: Fano and Breit-Wigner resonances}
\item{$\hbar\Omega<U$, high contact doping; with II.a above: inelastic resonant tunneling}
\end{enumerate}
\item{$Z_1>1$:} Strong drive
\begin{enumerate}
\item{$\hbar\Omega\gtrsim U$, low contact doping; with I.a above: multiple Fano and Breit-Wigner resonances}
\item{$\hbar\Omega<U$, high contact doping; with II.a above: inelastic resonant tunneling and high-harmonic generation}
\end{enumerate}
\end{enumerate}

We can estimate from experiments the typical parameter values.
Contact doping (parameter $U$) has been reported\cite{Giovannetti:2008dy,Laitinen:2016ht} in the range of -100 to 100 meV
(corresponding to doping levels of up to $10^{12}$ cm$^{-2}$, either $n$ or $p$-type).
Typical device channel lengths are from 10~nm to 1 $\mu$m,
making the corresponding energy scale $\hbar v_f/L$  in the range of $1-100$~meV.
The corresponding ballistic flight time from source to drain is $\tau=L/v_F$ and is about 1~ps.
We note in passing that within Landauer-B\"uttiker scattering theory, all relaxation times must then be longer than this,
which is the case at low temperature and low energies in a ballistic device (mobility $\mu\geq 10^5$ cm$^2$/Vs).
The driving frequency, $\hbar\Omega$, is between 0.4-40~meV for the THz frequency range $0.1-10$~THz.
The drive strength $Z_1$, for $Z_1\sim 1$, corresponds to a voltage of the order of a meV on the top gate for typical gate lengths
(see the estimate in our previous paper\cite{Korniyenko:2016ct}).
Finally, in the following we assume that temperature is the smallest energy scale (we put $T=0$).
With these numbers, all parameter regimes listed above are within experimental reach.

\begin{figure}[t]
\includegraphics[width=\columnwidth]{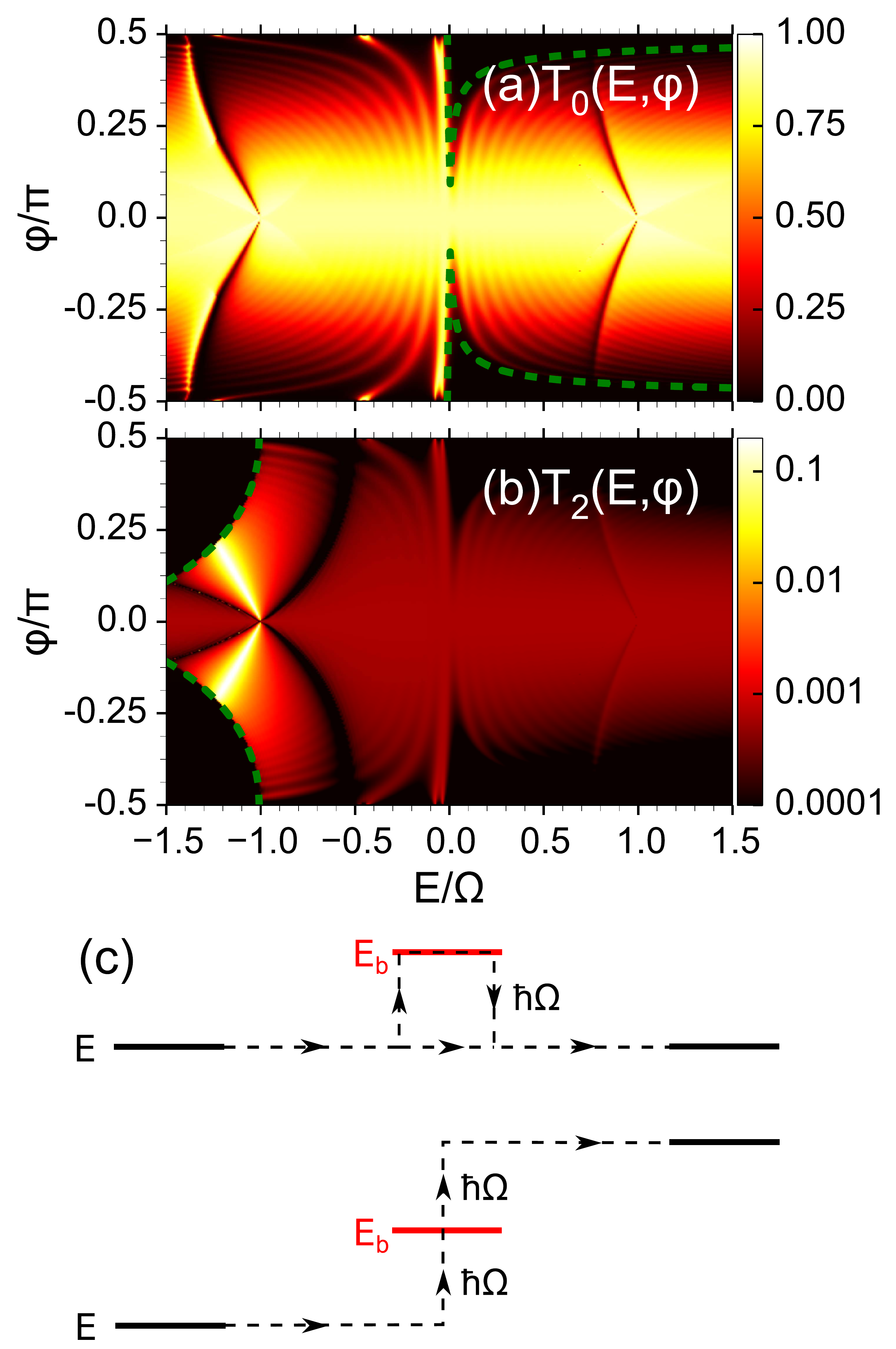}
\caption{(a) Direct transmission probability $T_0(E,\varphi)$ and (b) transmission probability to the second side band $T_2(E,\varphi)$
for parameters corresponding to Fig.~2 in Ref.~\onlinecite{Korniyenko:2016ct} ($Z_0=0.4\pi$, $Z_1=0.45$), but including
a small doping of contact leads $U=-0.01\Omega$ relative to the channel ($U_C=0$).
The device is long, such that $\hbar v_F/L=0.01\Omega$.
(c) Sketches of the Fano resonance process and the inelastic Breit-Wigner resonance process identified in Ref.~\onlinecite{Korniyenko:2016ct}
to be responsible for the dip-peak structure in $T_0$ and the peak in $T_2$, respectively.}
\label{fig:Fano_modified}
\end{figure}

\section{Weak to intermediate drive, $Z_1<1$}\label{sec:weak_drive}

\subsection{Low contact doping: Fano and Breit-Wigner resonances}

In Ref.~\onlinecite{Korniyenko:2016ct} we studied the case when $\hbar v_F/L$ is the smallest energy scale, i.e the channel is long.
We were then allowed to assume that evanescent waves can not reach between the contacts and the delta barrier.
In practice we set $U(x)=0$, and let $L\rightarrow\infty$.
In these limits, we studied Fano and Breit-Wigner resonances induced by the delta barrier quasibound state
and argued that they can be used to enhance the second harmonic.
In the more general formalism introduced here, we can ask the question what a small amount of contact doping $U\neq 0$ leads to.
We present in Fig.~\ref{fig:Fano_modified}(a)-(b) the transmission probabilities $T_0(E,\varphi)$ and $T_2(E,\varphi)$
for a small amount of contact doping and large distance to contacts, $|U|=\hbar v_F/L=0.01\Omega$.
Compared with the results in Ref.~\onlinecite{Korniyenko:2016ct} we find a small wedge of evanescent wave transport
in an energy window around $E=0$ (outside the green dashed lines).
The transmission of propagating waves displays fast Fabry-P\'erot interferences.
The Fano resonance in $T_0$ and the Breit-Wigner resonance in $T_2$ [processes sketched in Fig.~\ref{fig:Fano_modified}(c)] are however not affected.
For increasing contact doping (larger $|U|$), the Fabry-P\'erot oscillations get stronger
and will eventually interfere with the Fano and Breit-Wigner resonances, but not destroy them.
This holds as long as $\hbar v_F/L\ll\Omega$. For shorter contacts, the wedge of evenescent wave transport around $E=0$ widens.
When $\hbar v_F/L$ and $\Omega$ are of comparable magnitude, the most important feature in the transmission
is instead resonant inelastic tunneling.

\subsection{High contact doping: inelastic resonant tunnelling}

\begin{figure*}[t]
\includegraphics[width=\textwidth]{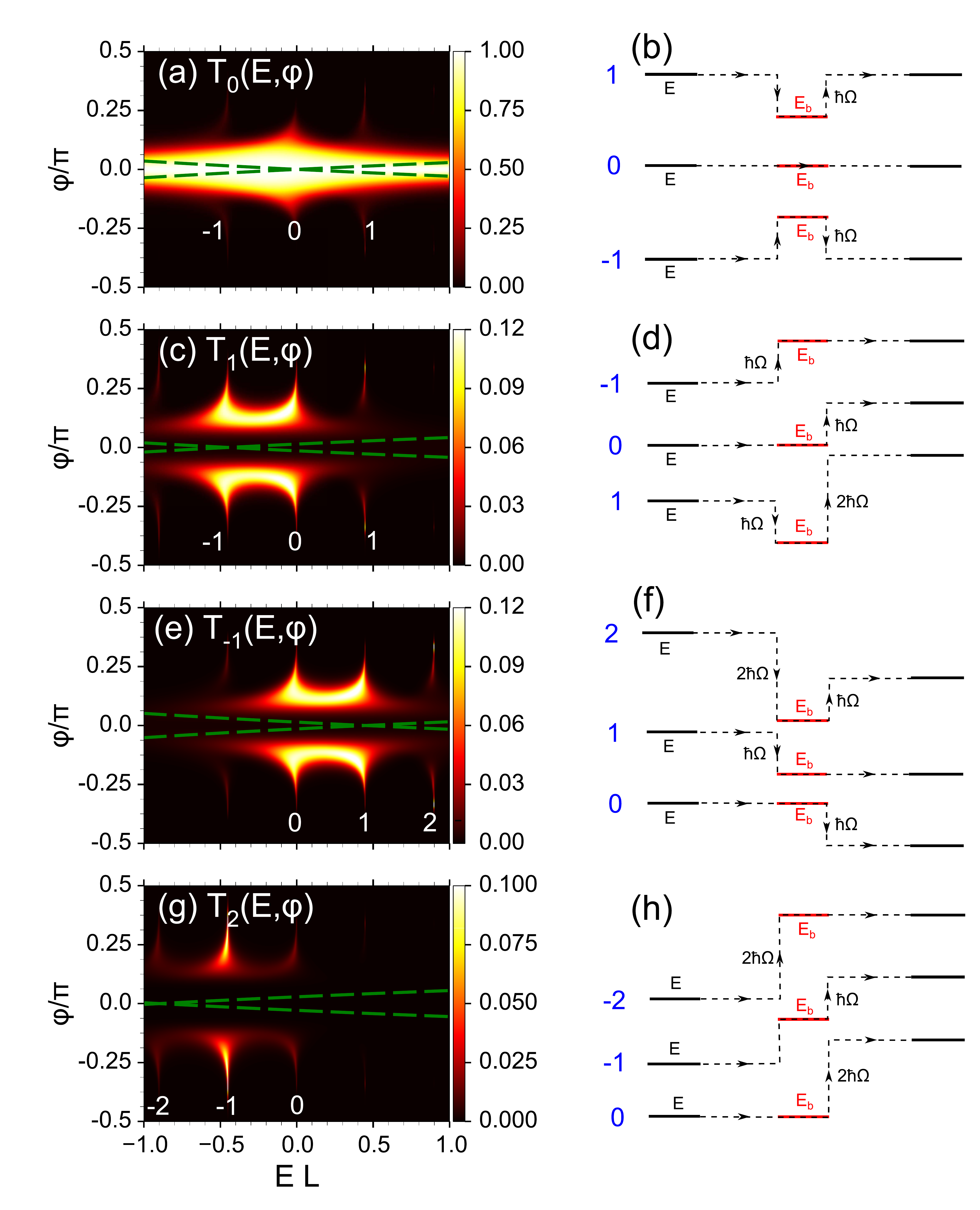}
\caption{Resonant transmission via evanescent waves for elastic transmission ($n=0$)
and inelastic transmission to sideband energies ($n=\pm 1$ and $n=2$).
The parameters are $U=-10/L$, $\Omega=0.45/L$, $Z_0=0.5\pi$, and $Z_1=0.1$.}
\label{fig:sidebands} 
\end{figure*}

\begin{figure}[t]
\includegraphics[width=\columnwidth]{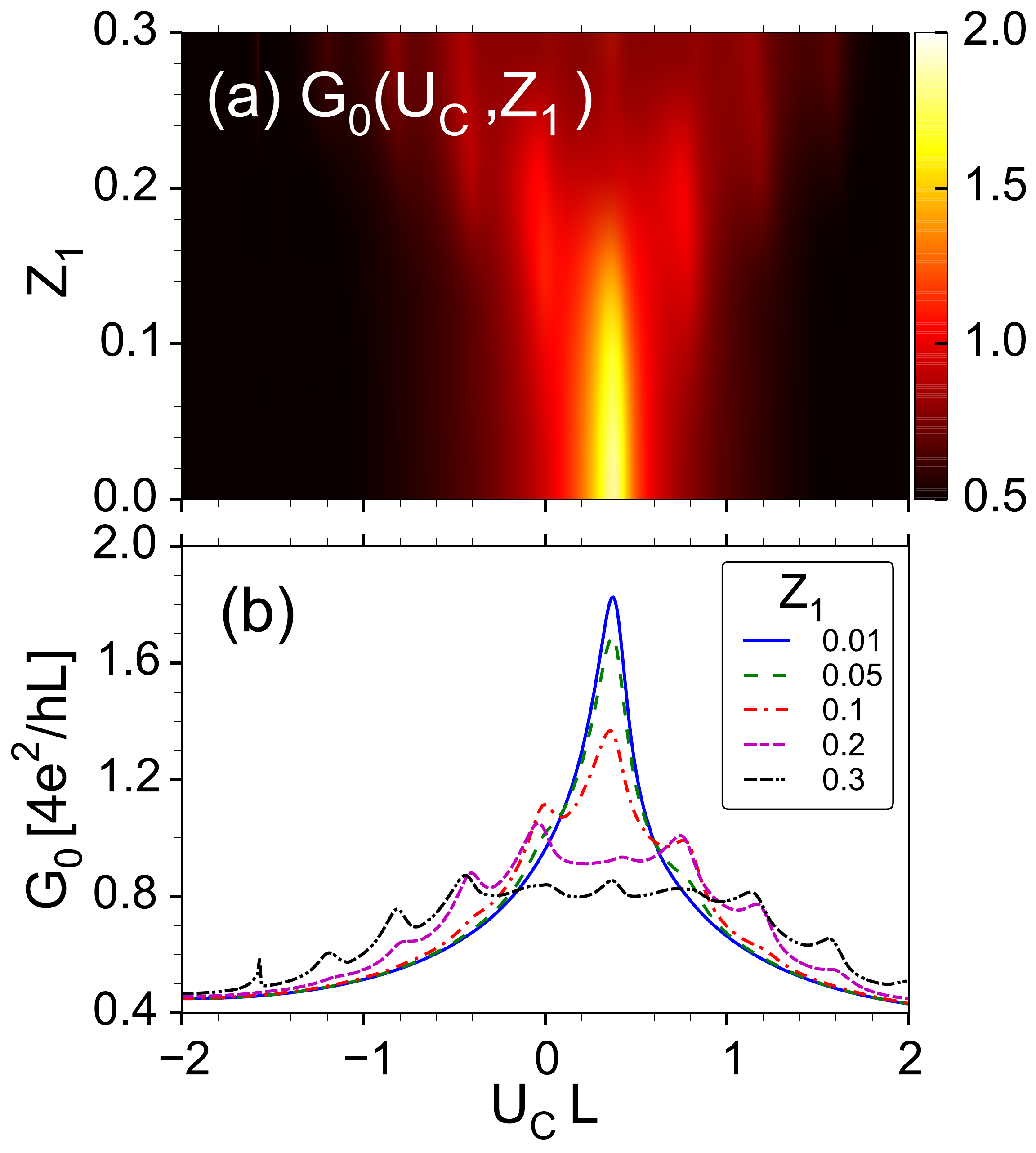}
\caption{\label{fig:g0map_small}(a) Dc conductance under ac drive of varying strength $Z_1$
in a range of channel dopings $U_C$ correspondning to evanescent wave transport.
The resonance peak for dc is reduced under ac drive and side peaks spaced by multiples of $\Omega$
appear due to resonant inelastic tunneling.
(b) Horizontal cuts in the colormap in (a) at particular values of $Z_1$.
The parameters are $Z_0=0.48\pi$, $U=-10/L$, and $\Omega=0.4/L$.}
\label{fig:G0_ac}
\end{figure}

Let us next consider the resonant tunneling regime. We assume a symmetric device with $L_1=L_2$,
with highly doped leads, and weakly doped channel, $|U_C|\ll |U|$, such that we have evanescent wave transport through the device.
The resonance due to the quasibound state in the delta barrier studied for dc transport above
will also create resonant inelastic tunneling under ac drive.
The resonance condition for weak drive $Z_1\ll 1$ is $n\Omega=E_b$.
This leads to promotion of higher-order sidebands as well as higher harmonics in the conductance that we will study below.

In Fig.~\ref{fig:sidebands} we present the transmission probabilities $T_n$ for $n=0$, $\pm 1$, and $2$.
For $T_0$ in Fig.~\ref{fig:sidebands}(a) two new transmission peaks emerge,
separated by $\pm\Omega$ from the main (0th) peak present already in dc.
The side peaks emerge because of possibility of absorbing/emitting energy quanta, as shown in panel (b) of the figure.
In the evanescent region, multiple sideband energies can now satisfy the bound state requirement,
thus resulting in a number of resonant peaks separated roughly by $\Omega$ (for $Z_0\approx \pi/2$).
Generally, these peaks are weaker than the one in the static case,
since the bound state contribution is now spread across several channels.
Analogous processes are involved during inelastic scattering between sidebands,
as illustrated in Fig.~\ref{fig:sidebands}(c) and (d) for $T_1$,
Fig.~\ref{fig:sidebands}(e) and (f) for $T_{-1}$,
and Fig.~\ref{fig:sidebands}(g) and (h) for $T_2$.

In Fig.~\ref{fig:G0_ac} we present the dc conductance $G_0$ as a function of channel doping $U_C$ for increasing ac drive strength $Z_1$.
The inelastic resonant tunneling processes discussed above for transmission probabilities result
in side peaks in the conductance spaced by multiples of $\Omega$ from the main resonance peak present in dc.
Already for rather weak drive $Z_0\sim 0.1$, several peaks are visible and the main elastic peak is reduced.
This can be traced to the energy dependence in the matrix on the left-hand side in Eq.~(\ref{t_n}),
which is given by a combination of functions in Eq.~(\ref{velocity}) all inversely proportional to energy.
The bound state energy $|E_b-U_C|$ is small, which results in division of small numbers and enhanced
effective coupling of sidebands close to the resonance energy. Thus, the range of validity of a perturbative
approach in small $Z_1$ is limited.

\begin{figure}[t]
\includegraphics[width=\columnwidth]{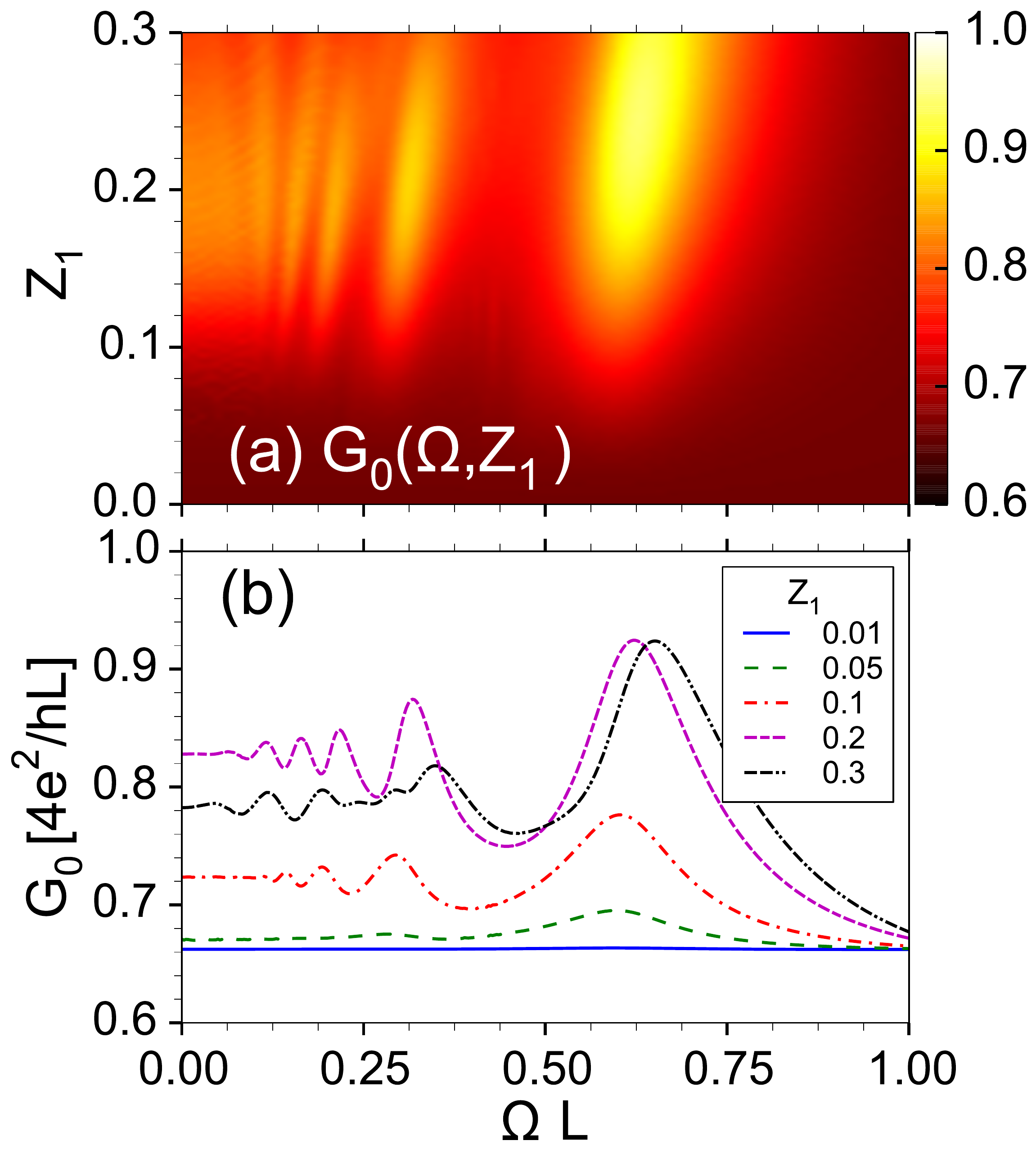}
\caption{(a) Dc conductance under varying ac drive frequency $\Omega$ for the same parameters
as in Fig.~\ref{fig:G0_ac}, but with fixed channel doping $U_C=1/L$.
Inelastic tunneling resonance peaks appear when a sideband coincides with the quasibound state.}
\label{fig:G0_ac_omega}
\end{figure}

In the literature, when other systems than graphene have been studied,
the conductance is often presented as a function of ac drive frequency \cite{2005PhR...406..379K}.
That is natural since there is often no knob corresponding to the very convenient back gate which can be used to tune
the graphene channel doping level (i.e. the parameter $U_C$ varied above).
For comparison, we present in Fig.~\ref{fig:G0_ac_omega} the dc conductance for varying frequency, keeping
the doping level $U_C=1/L$, i.e. a hole doped channel. In this case, we find conductance peaks at frequencies such
that a sideband coincides with the quasibound state, i.e. $n\Omega=E_b$.
Higher order processes are weaker for weak drive strength $Z_1$, thus the resonance peaks have smaller amplitudes and widths.

Considering Figs.~\ref{fig:G0_ac}-\ref{fig:G0_ac_omega} together, it is clear that the device can be used as a tunable detector.
The frequency $\Omega$ of the signal that needs to be detected tells us which channel doping we should choose
(tunable by the back gate), such that the first sideband is resonant.
Then the dc conductance is monitored to detect the signal.

\section{Strong drive, $Z_1>1$}\label{sec:strong_drive}

\begin{figure}
\subfigure{\includegraphics[width=\columnwidth]{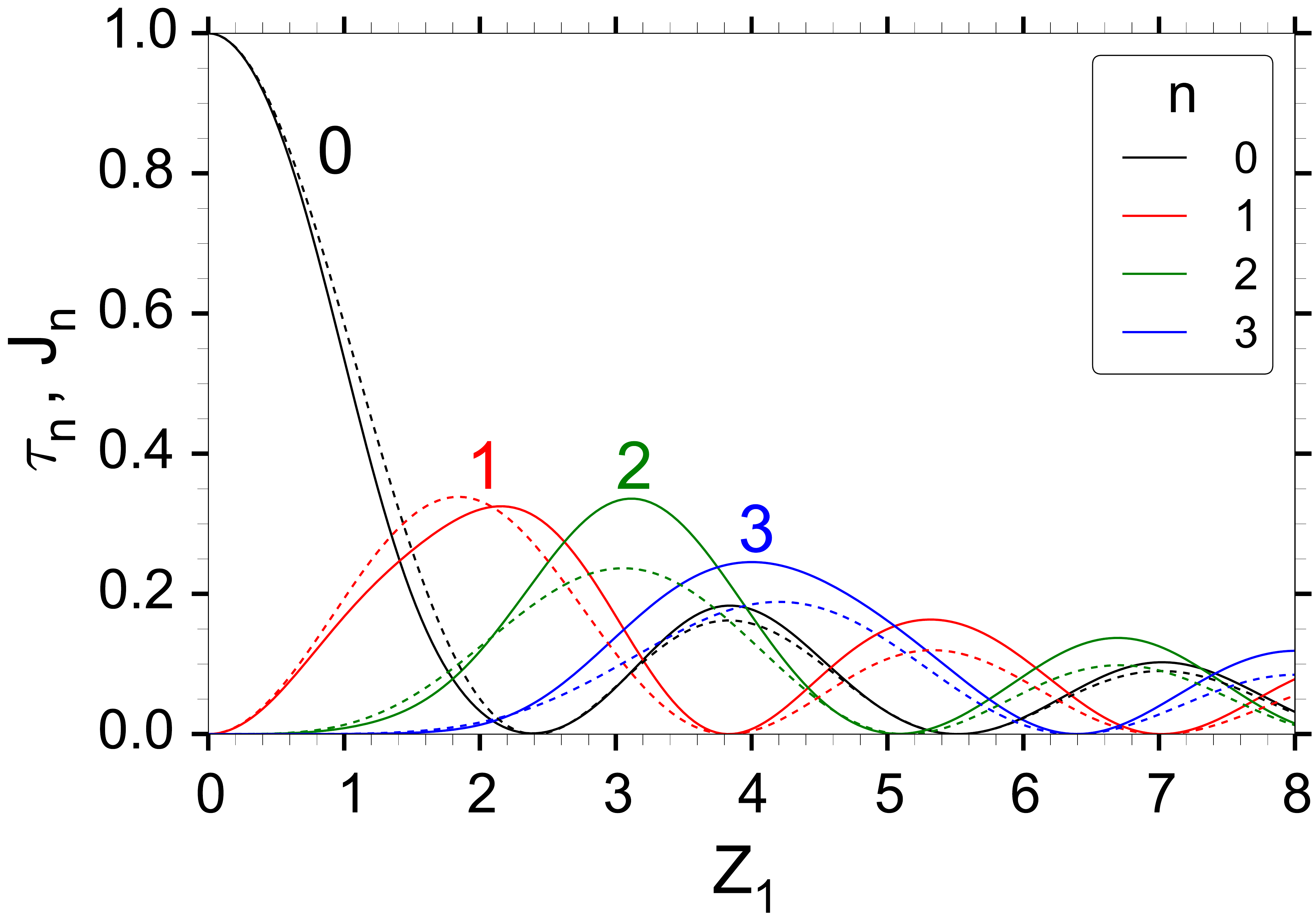}}
\caption{\label{fig:bessel} Bessel functions envelopes (dashed lines) and normalized angle-integrated sideband transmissions $T_n$ (solid lines)
for $U_C=-9/L$. The parameters are $U=-10/L$, $Z_0=0.4\pi$, and $\Omega=1/L$.}
\end{figure}

To understand the system behavior at strong drive it is useful to look at the transmission probability behavior as a function of driving strength $Z_1$. Since the $\delta$ barrier boundary condition matrix $\check M$ is directly related to Bessel functions of the first kind in sideband space, see Eq.~(\ref{eq:M_bessel}), we can expect transmission amplitudes to also follow corresponding Bessel functions. To illustrate the point, we introduce normalized angle-integrated transmissions
\be
\tau_n(U_C,Z_1) = \frac{ \int d\varphi\, T_n(\varphi,U_C, Z_1) }{ \int d\varphi\, T_0(\varphi,U_C, Z_1=0) }.
\ee
Indeed, the general behavior of $\tau_n$ for constant $U_C$ follows that of $J_n^2(Z_1)$, see Fig.~\ref{fig:bessel}.
Next, let us discuss how the resonances described above for weak drive evolve for strong drive, bearing in mind that
the distribution of sideband amplitudes is in simplified terms given by Bessel functions.

\subsection{Low contact doping}

\begin{figure}[t]
\includegraphics[width=\columnwidth]{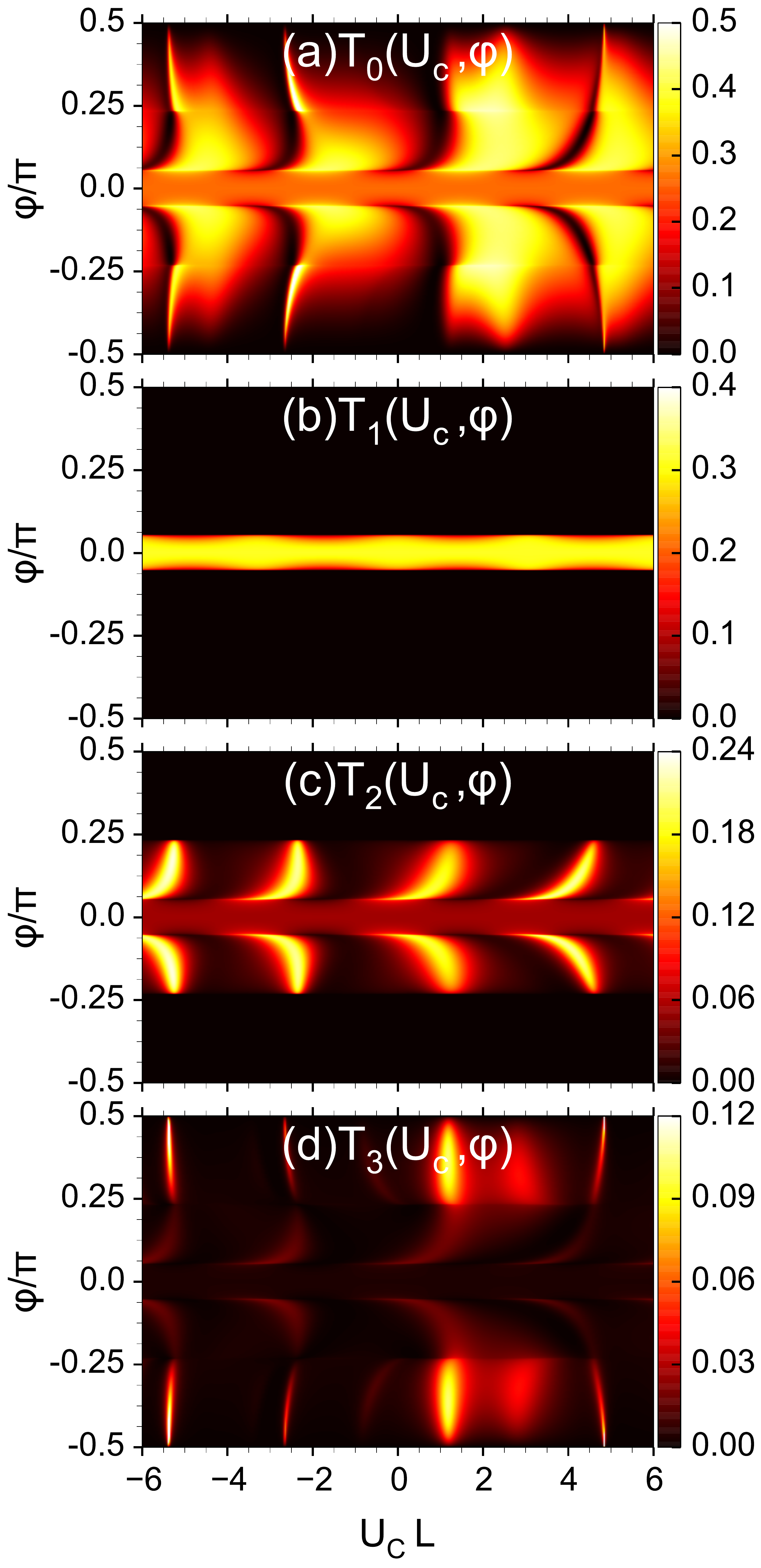}
\caption{Transmission functions $T_n(U_C,\varphi)$ for strong drive $Z_1=1.5$ and low contact doping $U=1.2/L$.
The drive frequency is $\Omega=1/L$, and the static barrier strength is $Z_0=0.4\pi$.}
\label{fig:transition}
\end{figure}

\begin{figure}[t]
\includegraphics[width=\columnwidth]{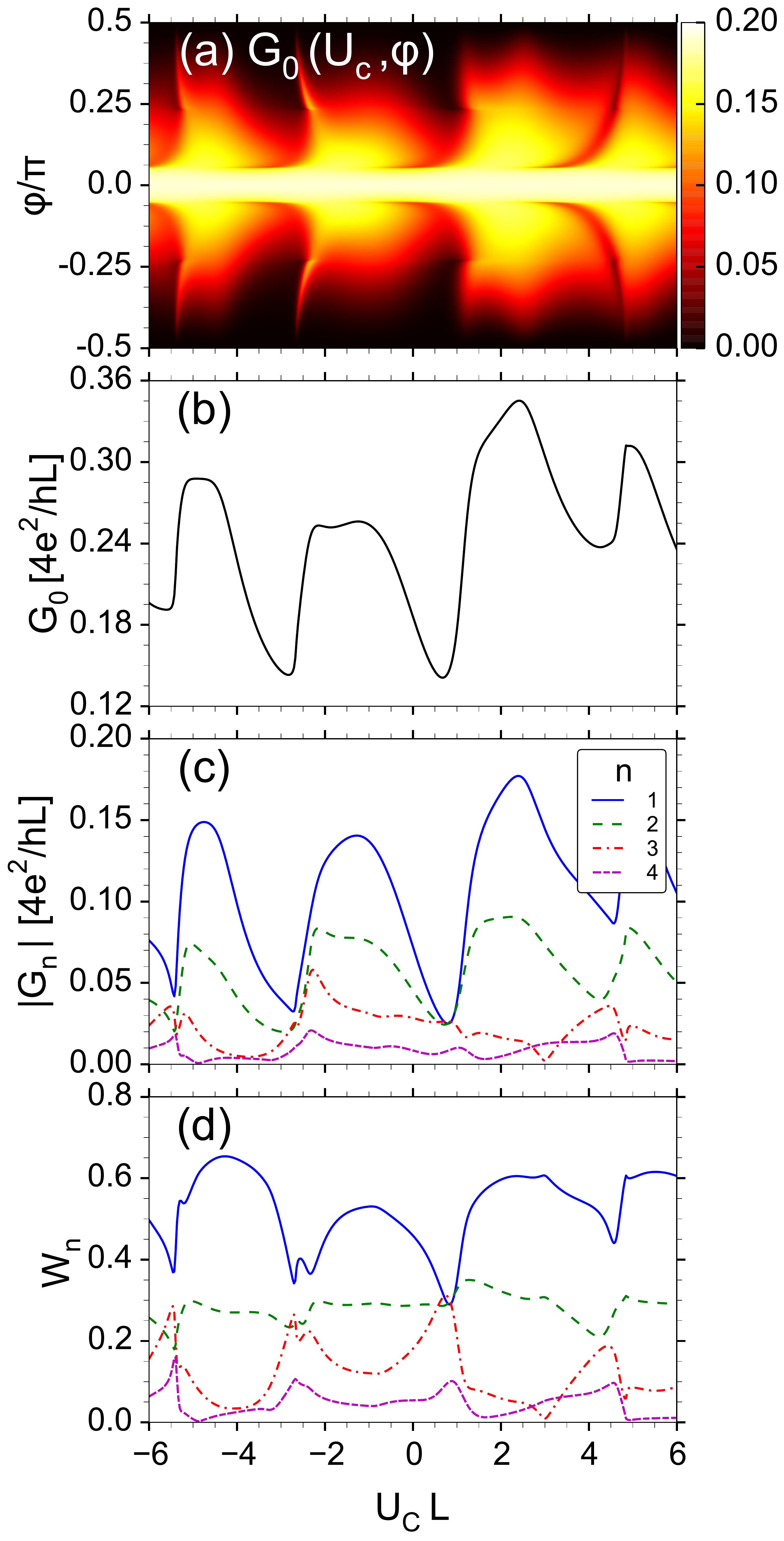}
\caption{(a) Angle resolved and (b) angle-integrated dc conductance for the same parameters as in Fig.~\ref{fig:transition}.
(c) First four ac harmonics, and (d) the relative weight of ac harmonics as defined in Eq.~(\ref{eq:Wn}).}
\label{fig:transitionGn}
\end{figure}

First, we would like to discuss the evolution of Fano and Breit-Wigner resonances described above for low doping $U$ of contacts. We observe multiple Fano resonances, c.f. Fig.~\ref{fig:transition}, that are due to the bound state condition satisfied by sideband waves in the contacts. It is useful to note, that since we fixed the energy $E_F=0$ and have $U_C$ as our parameter, the evanescent wave region boundaries for sideband $n$ become horizontal lines given by $|\varphi|=\phi_c^n$ where
\be
\phi_c^n=\arcsin\left|\frac{n\Omega-U}{U}\right|.
\label{eq:phi_c_n}
\ee
Note that this equation holds for $|n\Omega-U|<|U|$.
Otherwise waves are propagating in the contacts for all $\varphi$ and we can set $\phi_c^n=\pi/2$.
To avoid confusion, we emphasize that $\phi_c^n$ in Eq.~(\ref{eq:phi_c_n}) defines critical angles for evanescent sideband waves
in the contacts (which do not contribute to transport),
while $\varphi_c$ in Eq.~(\ref{eq:phi_c}) defines a critical angle for evanescent waves in the channel (which do contribute to transport).
For parameters used in Fig.~\ref{fig:transition}, only $n=1$ and $2$ sidebands have evanescent regions. Corresponding Fano and Breit-Wigner resonances now originate at the critical angle boundary and disperse with the angle of incidence.  As has been shown in our previous work \cite{Korniyenko:2016ct}, Fano resonances broaden as $Z_1^2$ and their positions change as the driving strength is increased. We note also that due to the strong coupling between sidebands for $Z_1>1$, the evanescent region boundary is clearly visible across all transmission channels. Unlike in the weak driving case, the zeroth transmission channel stops being dominant and thus higher sidebands are increasingly important in the conductance calculation.

Given the strong separation between evanescent and propagating wave regions as a function of angles in transmissions, it leads to a similar pronounced behaviour in angle resolved conductances, as shown in Fig.~\ref{fig:transitionGn}(a) for the dc component. After integration over angles, we observe clear oscillations in the $U_C$-dependence of the dc conductance, see Fig.~\ref{fig:transitionGn}(b), which are due to the multiple Fano resonances discussed above. The second and third harmonics are of equal size as the first harmonic for $U_c$ corresponding to the resonances, see Fig.~\ref{fig:transitionGn}(c).

It is useful to define a quantitative estimate of the relative power of ac harmonics as
\be
W_n=\frac{|G_n|}{\sum\limits_{n=1}^{\infty}|G_n|}, \quad n\geq 1.
\label{eq:Wn}
\ee
For simplicity we exclude negative $n$ harmonics in this estimate, since we know that $G_{-n}=G_n^*$.
In Ref.~\onlinecite{Korniyenko:2016ct} we discussed weak drive and second harmonic generation.
In Fig.~\ref{fig:transitionGn}(d) we show for strong drive $Z_1=1.5$ that both second and third harmonic can be
resonantly enhanced and become of the same order as the first harmonic for the case $|U_C|>|U|$.
Higher harmonics $n>3$ are however not enhanced above the first harmonic
in the regime of low contact doping $U$, even for stronger $Z_1$, because the multiple resonances are not equidistant in energy space.

\subsection{High contact doping}

\begin{figure}[t]
\includegraphics[width=\columnwidth]{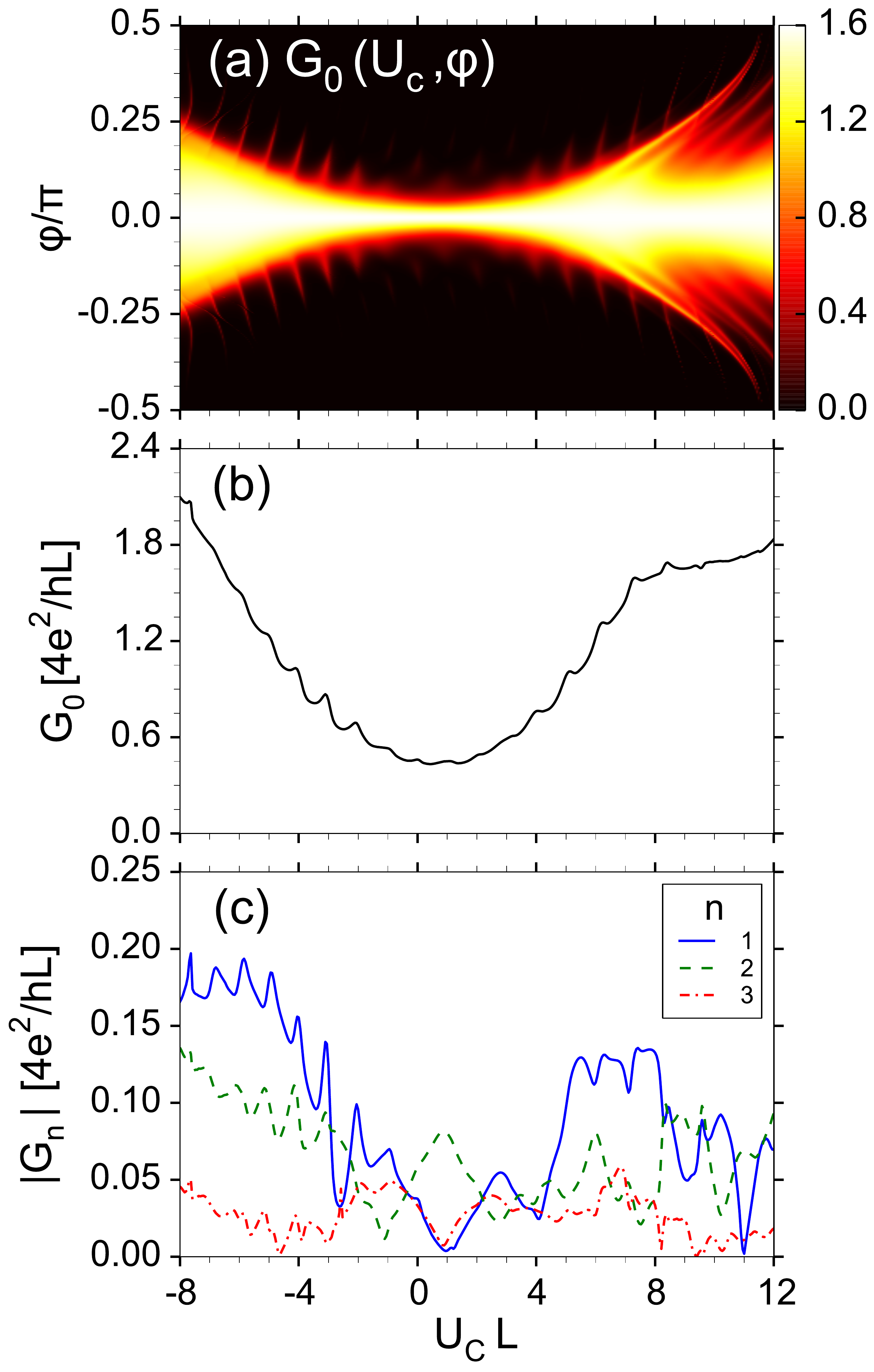}
\caption{(a) Angle resolved, and (b) angle-integrated dc conductance for high contact doping $U=-10/L$ and strong drive $Z_1=1.5$.
(c) First few ac harmonics for the same parameters. The drive frequency is $\Omega=1/L$ and the static barrier strength is $Z_0=0.4\pi$.}
\label{fig:highU}
\end{figure}

Next, let us study the effect of strong contact doping ($|U|$ large).
See Fig.~\ref{fig:highU}(a) for the angular dependence of the dc conductance in this case.
A clear valley in the dc conductance is centered at the Dirac point in the central region (i.e. around $U_C=0$),
which corresponds to evanescent wave transport in the channel.
The double-barrier inelastic tunneling resonances at $n\Omega=E_b$ result in a fine comb of equidistant peaks inside the valley.
After integration over angles, see Fig.~\ref{fig:highU}(b), the dc conductance shows small oscillations related to the inelastic tunneling processes.
Note that the weight of the resonance peak we studied in the absence of ac drive ($Z_1=0$) in Fig.~\ref{fig:G0}(b), has been completely redistributed
across the many peaks of the comb.
The peak period ($\Omega$) is the same for all transmission channels.
Therefore, analogical fine oscillations show up in ac harmonics as well, see Fig.~\ref{fig:highU}(c).

\begin{figure}[t]
\includegraphics[width=\columnwidth]{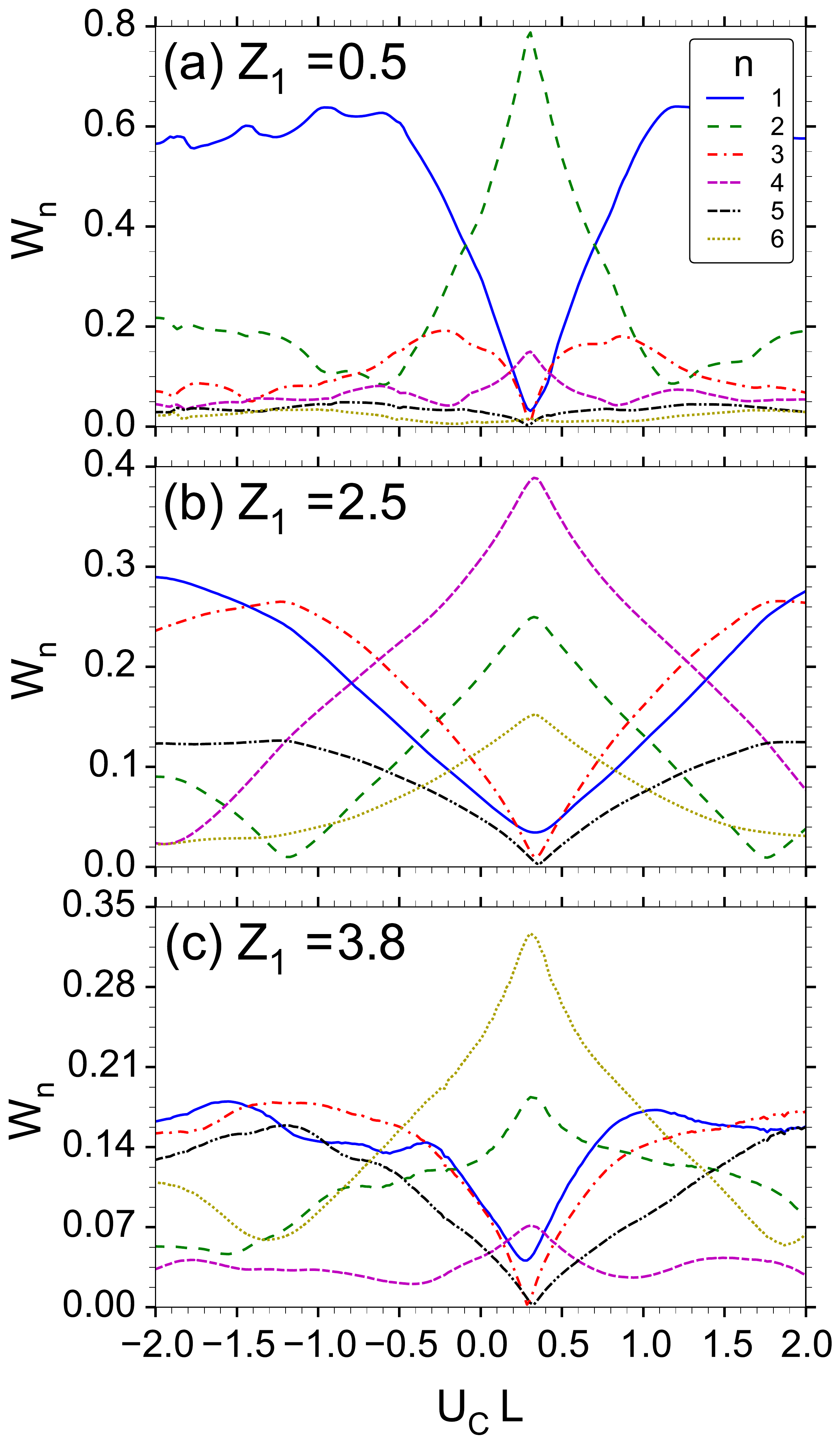}
\caption{High harmonic enhancement for (a) $Z_1=0.5$, (b) $Z_1=2.5$, and (c) $Z_1=3.8$.
The parameters are $U=-10/L$, $Z_0=0.48\pi$, and $\Omega=0.45/L$.}
\label{fig:proposal1}
\end{figure}

\begin{figure}[t]
\includegraphics[width=\columnwidth]{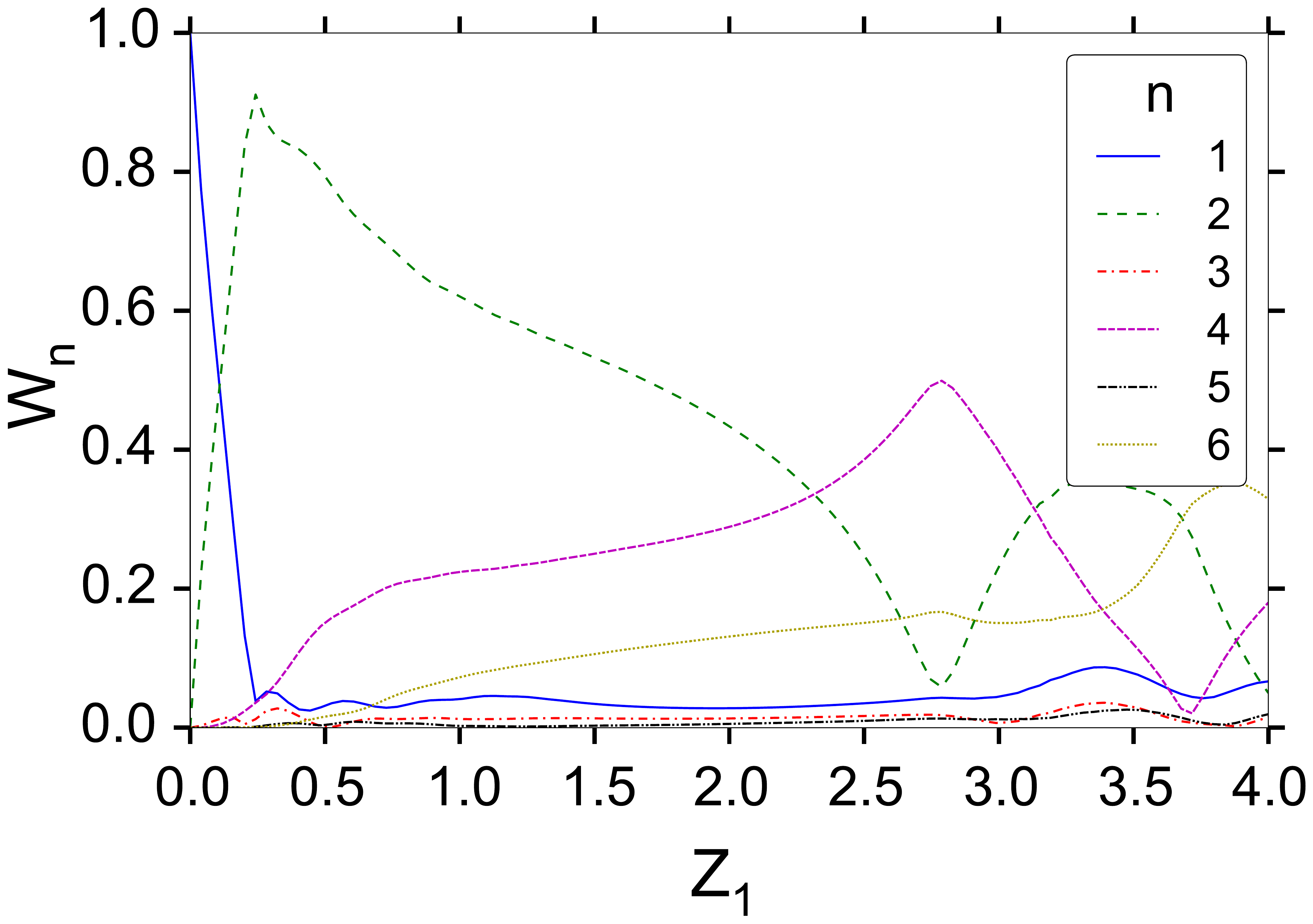}
\caption{High harmonic enhancement with increasing drive strength $Z_1$ for $U_C=0.3/L$, corresponding to on-resonance transport. The parameters are the same as in Fig.~\ref{fig:proposal1}.}
\label{fig:Wn}
\end{figure}

In Fig.~\ref{fig:highU}(c) we note that for $U_C$ corresponding to the direct double-barrier resonance,
the second harmonic is enhanced above the first harmonic, which is suppressed.
By tuning parameters, we can in fact enhance a selected even $n$ harmonic, as shown in Fig.~\ref{fig:proposal1},
where we present the weights $W_n$ as a function of channel doping $U_C$ for increasing drive strength $Z_1$.
In panels (a)-(c) we obtain the $n=2$, $n=4$, and $n=6$ harmonic, respectively.
To emphasize this result, we plot the distribution between harmonics roughly on resonance ($U_C=0.3/L$) as a function of $Z_1$ in Fig.~\ref{fig:Wn}.
In the whole range of drive strengths $Z_1>0.25$, all odd $n$ harmonics are suppressed, while even $n$ harmonics are enhanced, one after the other.

For weak drive strength $Z_1$, we can show that $G_1$ is suppressed while $G_2$ is enhanced through
destructive (for $G_1$) and constructive (for $G_2$) interferences between the transmission processes responsible
for the corresponding harmonic, c.f. Eq.~(\ref{eq:conductance}).
To explain this behavior, we first note that Eq.~(\ref{eq:M_ac_bessel}) tells us that
sideband amplitudes $t_n$ are proportional to $i^{|n-m|}J_{|n-m|}(Z_1)$ to lowest order in $Z_1$.
For instance $t_0\propto J_0$ and $t_{\pm 1}\propto iJ_1$ (symmetric).
It follows that for small $Z_1$ the first conductance harmonic (before integration over transverse momentum) can be written as
a sum of two terms involving two different transfer processes:
\begin{eqnarray}
G_1(E,\varphi) &\propto& t_{-1}^*(E,\varphi)t_0(E,\varphi) + t_0^*(E,\varphi)t_1(E,\varphi)\nonumber\\
&=& J_0J_1\left[-ic_{-1}(E,\varphi)+ic_1(E,\varphi)\right],
\end{eqnarray}
where $c_{\pm 1}$ are complex numbers.
On resonance, $c_{-1}\approx c_{1}$, the two terms cancel, and $G_1$ is suppressed.
That this symmetry appears on resonance can be seen from Fig.~\ref{fig:sidebands}, where the peaks in $T_{\pm 1}(E_b)$ [panels (c) and (e)],
corresponding to processes labeled 0 [panels (d) and (f)], have the same shape and magnitude. Off resonance, the probabilities $T_{\pm 1}(E)$
are obviously not equal, the two terms do not cancel, and $G_1$ is not suppressed.
For the enhancement of the second harmonic, we note that for small $Z_1$ we have $G_2\propto t_{-2}^*t_0 + t_{-1}^*t_1 + t_{0}^*t_2$.
All terms consist of real products of the coupling matrix elements $M_{nm}$ in sideband space.
On resonance, for instance $t_{-2}^*t_0+t_0^*t_2\propto J_0J_2(c_{-2}+c_2)$ with $c_{-2}\approx c_2$,
and the two terms sum up constructively because of the real coupling in sideband space.
For stronger drive, the odd (even) harmonics are suppressed (enhanced) in an analogous way,
where pairs of processes add up destructively (constructively).

\section{Summary}\label{sec:summary}

We have presented results for the ac conductance in a ballistic graphene field-effect transistor with a time-modulated top gate potential,
including an inhomogeneous doping profile across the device.
We have studied two regimes, corresponding to (i) low doping of contacts and (ii) high doping of contacts, relative
to the doping level in the channel (which is tunable by a back gate).
For case (i) we find Fano resonances in direct transmission and Breit-Wigner resonances in inelastic scattering to sideband energies.
The resonances are due to excitation of quasibound states in the channel, analogous to what we found in Ref.~\onlinecite{Korniyenko:2016ct}.
Here we have shown that these resonances survive when a moderately varying doping landscape across the device is taken into account.
For case (ii) we find inelastic tunneling resonances via quasibound states in the top gate barrier potential.
For weak drive, the resonances lead to a large response in the direct current between source and drain already for weak ac drive on the top gate.
We propose that the device can be utilized as a detector in the THz frequency range.
In addition, for strong drive, inelastic tunneling to multiple sidebands results in resonant excitation of higher harmonics $n\Omega$
[we demonstrate dominance of $n=6$ in Fig.~\ref{fig:proposal1}(c)],
with $n$ an even number due to an interference effect between different tunneling processes.
The harmonic $n$ (even) can be selected either by the back gate or by tuning the drive strength.
In summary, ac transport in ballistic graphene field-effect transistors is a rich subject for studying quantum mechanical
resonance phenomena that can possibly also be utilized in applications such as detectors of THz radiation or
to generate high harmonics.

\acknowledgments

We acknowledge financial support from the Swedish foundation for strategic research (SSF), the Knut and Alice Wallenberg foundation (KAW), and the Swedish research council. The research of OS was partly supported by the National Science Foundation (Grant DMR-1508730).

\begin{appendix}

\section{Wave solutions - static case}\label{sec:appendix_static}

First we derive the wave solutions to the Dirac equation without time-dependent perturbation.
We assume translational invariance and conserved parallel momentum $k_y$,
in which case the Hamiltonian has the form
\begin{equation}
\mathcal{H}_0 = -i\sigma_x\nabla_x + \sigma_y k_y + Z_0\delta(x) + U(x),
\label{GrapheneHam}
\end{equation}
where the device doping profile is described by
\begin{equation}
U(x) = U_L\theta(-L_1-x) + U_R\theta(x-L_2).
\end{equation}
This means that in this derivation we choose the Dirac point in the device channel, $x\in[-L_1,L_2]$, as reference level where $E=0$ (i.e. $U_C=0$).
The static Dirac equation
\begin{equation}
\mathcal{H}_0\psi(x,k_y,E)=E\psi(x,k_y,E)
\end{equation}
is straightforward to solve by making a plane-wave ansatz and find unknown coefficients through boundary conditions.
But first it is convenient to introduce a scattering basis.

\subsection{Scattering basis}

Consider the homogeneous case, i.e. $Z_0=0$ and $U(x)=0$ in Eq.~(\ref{GrapheneHam}).
The solutions, labeled by $k_y$ and $E$, can be organized into a scattering basis for right-
and left-moving (along the $x$-axis) plane waves, as defined by their group velocities.
This scattering basis has the form
\begin{align}
&\psi_{\rightarrow}(x,k_y,E) = \frac{1}{\sqrt{2\mathrm{v}(k_y,E)}}
\begin{pmatrix}
1\\
\eta(k_y,E)
\end{pmatrix}
e^{i\kappa(k_y,E)x},\notag\\
&\psi_{\leftarrow}(x,k_y,E) = \frac{1}{\sqrt{2\mathrm{v}(k_y,E)}}
\begin{pmatrix}
1\\
\bar{\eta}(k_y,E)
\end{pmatrix}
e^{-i\kappa(k_y,E)x},\label{ScatBasis}
\end{align}
where
\begin{align}
&\eta(k_y,E) = \frac{\kappa(k_y,E)+ik_y}{E},\notag\\
&\bar{\eta}(k_y,E) = \frac{-\kappa(k_y,E)+ik_y}{E},\notag\\
&\mathrm{v}(k_y,E) = \frac{\kappa(k_y,E)}{E},\label{velocity}\\
&\kappa(k_y,E) = \mathrm{sgn}(E)\sqrt{E^2-k_y^2}\notag
\end{align}
The normalization of these plane waves is such that they carry unit probability flux along the $x$-axis, defined as
\begin{equation}
j_x(x,k_y,E) = \psi^{\dagger}(x,k_y,E)\sigma_x\psi(x,k_y,E).
\end{equation}
That is, we have $j_x^{\rightarrow}=1$ and $j_x^{\leftarrow}=-1$.
This scattering basis is useful in deriving the scattering matrix and computing the current within the Landauer-B\"uttiker formalism,
as we described in detail in Ref.~\onlinecite{Korniyenko:2016ct} for the case $U(x)=0$.

\subsection{Scattering matrix derivation}

Below we solve the scattering problem for quasiparticles at energy $E$ injected from the left contact at
conserved transverse momentum $k_y$ given by
\be
k_y=|E-U_L|\sin\varphi,
\ee
where $\varphi$ is the incidence angle on the scattering region, measured with respect to the $x$-axis.
There are four regions in our device: left and right contacts labeled $L$ and $R$
and left and right channel regions (with respect to the delta potential barrier) labeled $1$ and $2$.
The scattering state ansatz si then
\begin{widetext}
\begin{align}
\psi(x,k_y,E) = 
\begin{cases}
              \psi_{\rightarrow}(x,k_y,E-U_L) + r(k_y,E) \psi_{\leftarrow}(x,k_y,E-U_L),  & x<-L_1,\\
a(k_y,E) \psi_{\rightarrow}(x,k_y,E)        + b(k_y,E) \psi_{\leftarrow}(x,k_y,E),         & -L_1<x<0\\
c(k_y,E) \psi_{\rightarrow}(x,k_y,E)        + d(k_y,E) \psi_{\leftarrow}(x,k_y,E),         & 0<x<L_2\\
t(k_y,E)  \psi_{\rightarrow}(x,k_y,E-U_R), & x>L_2.
\end{cases}\label{dc_ansatz}
\end{align}
\end{widetext}
Note that the doping level in the channel region ($U_C=0$) is different from that in the  contacts ($U_L$ and $U_R$).
As a consequence, the waves can be evanescent in the channel region. This is included in the ansatz above by allowing $\kappa(k_y,E)$
in Eq.~(\ref{velocity}) to be imaginary. The convention we use is that $\Psi^{\rightarrow}$ denotes a wave evanescent towards
positive $x$, while $\Psi^{\leftarrow}$ denotes a wave evanescent in the opposite direction. This means that if $U_R\neq U_L$
and $\kappa^R=\kappa(k_y,E-U_R)$ turns imaginary, the ansatz above also holds,
although in this case $t(k_y,E)$ is not a transmission amplitude. It is then eliminated in favor of the reflection
coefficient $r(k_y,E)$, with $|r(k_y,E)|=1$. This is not so important in the present discussion, but becomes important in the following section
on ac transport. In the main text we only consider the special case $U_R=U_L$ for simplicity.

The coefficients in Eq.~(\ref{dc_ansatz}) are found through the boundary conditions, which are simple wave continuity at 
$x=-L_1$ and $x=L_2$, and a pseudospin rotation operation at the delta barrier (c.f. Ref.~\onlinecite{Korniyenko:2016ct}):
\bea
\psi(-L_1^-,k_y,E) &=& \psi(-L_1^+,k_y,E),\label{bc1}\\
\psi(0^-,k_y,E)     &=& \exp[iZ_0\sigma_x]\psi(0^+,k_y,E),\label{bc2}\\
\psi(L_2^-,k_y,E) &=& \psi(L_2^+,k_y,E).\label{bc3}
\eea

From Eq.~(\ref{bc3}) we can obtain $c(k_y,E)$ and $d(k_y,E)$ in terms of $t(k_y,E)$
\bea
c &=& \sqrt{\frac{v}{v^R}} \frac{\eta^R-\bar\eta}{\eta-\bar\eta} e^{i(\kappa^R-\kappa)L_2} t\\
d &=& \sqrt{\frac{v}{v^R}} \frac{\eta-\eta^R}{\eta-\bar\eta} e^{i(\kappa^R+\kappa)L_2} t.
\eea
Note that $v^R=v(k_y,E-U_R)$, and analogous for $\kappa^R$ and $\eta^R$
(also, $v^L$, $\kappa^L$ etc. appearing below are computed at energy $E-U_L$).
The quantities in regions $1$ and $2$, computed at energy $E$, lack superscripts.
Above and in the following we suppress the explicit reference to the dependences on $k_y$ and $E$ unless necessary.

From Eq.~(\ref{bc2}) we then obtain $a$ and $b$ in terms of $t$
\bea
a &=& \frac{\sqrt{2v}}{\bar\eta-\eta} \hspin(\bar\eta,-1)\exp[iZ_0\sigma_x] \vec B t,\\
b &=& \frac{\sqrt{2v}}{\eta-\bar\eta} \hspin(\eta,-1)      \exp[iZ_0\sigma_x] \vec B t,
\eea
where
\begin{equation}
\vec B = \left[ \frac{\eta^R-\bar\eta}{\eta-\bar\eta} \spin(1,\eta) e^{-i\kappa L_2} 
                   + \frac{\eta-\eta^R}{\eta-\bar\eta} \spin(1,\bar\eta) e^{i\kappa L_2}
              \right] \frac{e^{i\kappa^R L_2}}{\sqrt{2v^R}}.
\end{equation}

Finally, from Eq.~(\ref{bc1}) we obtain reflection and transmission coefficients
\bea
r &=& \vec C^{\,T} \exp[iZ_0\sigma_x] \vec B t,\label{r_static}\\
t &=& \left( \vec A^{\,T} \exp[iZ_0\sigma_x] \vec B \right)^{-1},\label{t_static}
\eea
where
\begin{equation}
\vec A = \left[ \frac{\bar\eta^L-\eta}{\eta-\bar\eta} \spin(\bar\eta,-1) e^{-i\kappa L_1} 
                   + \frac{\bar\eta-\bar\eta^L}{\eta-\bar\eta} \spin(\eta,-1) e^{i\kappa L_1}
              \right] \frac{e^{i\kappa^L L_1}}{\sqrt{2v^L}},
\end{equation}
and
\begin{equation}
\vec C = \left[ \frac{\eta-\eta^L}{\eta-\bar\eta} \spin(\bar\eta,-1) e^{-i\kappa L_1} 
                   + \frac{\eta^L-\bar\eta}{\eta-\bar\eta} \spin(\eta,-1) e^{i\kappa L_1}
              \right] \frac{e^{-i\kappa^L L_1}}{\sqrt{2v^L}}.
\end{equation}
The superscript $T$ in Eqs.~(\ref{r_static})-(\ref{t_static}) denotes transposition.

\subsection{Double barrier tunneling}\label{Appendix:resonant_tunneling}

Since waves are always propagating inside the delta potential, the channel regions on either side of it form a double tunnel barrier
when lead regions are highly doped such that waves are propagating there as well.
It is well-known that the bound state in this structure can lead to resonances in the transmission amplitude derived above.
To understand it qualitatively we write down a propagation matrix that relates amplitudes $a$ and $b$ 
at the left edge of the channel to amplitudes $c$ and $d$ at the right edge, see Fig.~\ref{fig:contacts}, i.e.
\be
\label{eq:pb3}
\spin(a,b) = P_b \spin(c,d)
\ee
where
\be
\label{eq:pb}
P_{b}=
\left(
\ba{cc}
e^{-i\kappa L_1}& 0\\
0 & e^{i\kappa L_1}
\ea
\right)
\hat D
\left(
\ba{cc}
e^{-i\kappa L_2}& 0\\
0 & e^{i\kappa L_2}
\ea
\right).
\ee
The four elements of the $2\times 2$-matrix $\hat D$ are obtained from the boundary condition at the delta barrier Eq.~(\ref{bc2}) as
\begin{align}
\label{eq:D}
D_{11}=\frac{1}{2v}\hspin(-\bar\eta,1)\exp[iZ_0\sigma_x]\spin(1,\eta),\\
D_{12}=\frac{1}{2v}\hspin(-\bar\eta,1)\exp[iZ_0\sigma_x]\spin(1,\bar\eta),\\
D_{21}=\frac{1}{2v}\hspin(\eta,-1)\exp[iZ_0\sigma_x]\spin(1,\eta),\\
D_{22}=\frac{1}{2v}\hspin(\eta,-1)\exp[iZ_0\sigma_x]\spin(1,\bar\eta).
\end{align}
For the case of evanescent waves in the channel, the wavevector becomes imaginary $\kappa=i\varkappa$
and Eq.~(\ref{eq:pb}) takes the form
\be
\label{eq:pb2}
P_{b}=
\left(
\ba{cc}
D_{11}e^{\varkappa L}& D_{12}e^{-\varkappa \Delta L}\\
D_{21}e^{\varkappa \Delta L} & D_{22}e^{-\varkappa L}
\ea
\right),
\ee
where $L=L_1+L_2$ and $\Delta L=L_1-L_2$.

For a symmetric system with $\Delta L=0$, and on resonance, i.e. when the energy $E$ of the scattering state
coincides with the delta barrier bound state $E_b$, it follows from the derivation in Ref.~\onlinecite{Korniyenko:2016ct} [c.f. Eq.~(B4)]
that $D_{11}=0$. When that happens, we see that Eq.~(\ref{eq:pb3}) with Eq~(\ref{eq:pb2}) leads to
\bea
a = D_{12}d,\label{eq:cross_connection}\\
b\approx D_{21} c,
\eea
where we also noted that $D_{22}\exp(-\kappa L)\ll D_{21}$.
This shows the cross connection between decaying and exploding solutions illustrated in Fig.~\ref{fig:contacts}(d).
When transmission is enhanced to unity, the exponential functions due to tunneling through the two barriers cancel each other.
Off resonance, this clean-cut cross connection does not occur and the transmission is exponentially suppressed.

\section{Wave solutions - dynamic case}\label{sec:appendix_dynamic}

Let us now derive the Floquet scattering matrix in presence of an oscillating delta barrier.
The Hamiltonian we consider is
\begin{equation}
\mathcal{H} = \mathcal{H}_0 + Z_1\cos(\Omega t)\delta(x).\label{FullHam}
\end{equation}
The time-dependent Dirac equation
\begin{equation}
\mathcal{H}\psi(x,k_y,t)=i\partial_t\psi(x,k_y,t),
\end{equation}
including a time-periodic  potential as in Eq.~(\ref{FullHam}), can be solved by making use of the Floquet ansatz:
\begin{equation}
\psi(x,k_y,t) = e^{-iEt}\sum_{n=-\infty}^{+\infty}e^{-in\Omega t}\psi_n(x,k_y,E).\label{Floquet_ansatz}
\end{equation}
In analogy with the static case above, this ansatz is made in each region.
Coefficients for transmitted and reflected waves are then contained in the amplitudes $\psi_n(x,k_y,E)$.
The coefficients are determined through the boundary conditions.
A complication in the dynamic case is the boundary condition at the oscillating delta barrier,
which mixes amplitudes at different sideband energies $E_n=E+n\Omega$.
Following Ref.~\onlinecite{Korniyenko:2016ct}, the boundary condition is best formulated by first introducing
a column vector with the many sideband amplitudes $\psi_n(x,k_y,E)$,
\begin{equation}
\Phi(x,k_y,E) = \left(
\begin{array}{c}
\dots\\
\psi_{-1}(x,k_y,E)\\
\psi_0(x,k_y,E)\\
\psi_{1}(x,k_y,E)\\
\dots
\end{array}\right).
\end{equation}
The condition to be satisfied at $x=0$ is then
\begin{align}
&\Phi(0^{-},k_y,E) = \check{M}\Phi(0^{+},k_y,E),\notag\\
&\check{M} = \exp\left[iZ_0\sigma_x\otimes\check\Gamma_0+i\frac{Z_1}{2}\sigma_x\otimes\check\Gamma_1\right],\label{expM}\\
&\left[\check\Gamma_0\right]_{n,m} = \delta_{n,m},\;\;\left[\check\Gamma_1\right]_{n,m} = \delta_{n,m+1}+\delta_{n,m-1}.\notag
\end{align}

The barrier scatters an incident wave labeled by $E$ and $k_y$ into a linear combination of waves labeled by $E_n$ and $k_y$.
In the end, when calculating transport properties, we have to consider only propagating outgoing waves in the leads,
$|E_n-U_L|>|k_y|$ and $|E_n-U_R|>|k_y|$. We use the following ansatz:
\begin{widetext}
\begin{align}
\psi_n(x,k_y,E) = 
\begin{cases}
\delta_{n0}\psi_{\rightarrow}(x,k_y,E_n-U_L)+r_n\psi_{\leftarrow}(x,k_y,E_n-U_L),& x<-L_1,\\
a_n\psi_{\rightarrow}(x,k_y,E_n)+b_n\psi_{\leftarrow}(x,k_y,E_n),& -L_1<x<0,\\
c_n\psi_{\rightarrow}(x,k_y,E_n)+d_n\psi_{\leftarrow}(x,k_y,E_n),& 0<x<L_2,\\
t_n\psi_{\rightarrow}(x,k_y,E_n-U_R),& x>L_2.
\end{cases}
\end{align}

The three boundary conditions can be written as
\bea
\psi_n(-L_1^-,k_y,E) &=& \psi_n(-L_1^+,k_y,E),\\
\psi_n(0^-,k_y,E)      &=& \sum_m \check{M}_{nm}\psi_m(0^+,k_y,E),\\
\psi_n(L_2^-,k_y,E)  &=& \psi_n(L_2^+,k_y,E).
\eea
The steps to solve for the coefficients are analogous to the static case and we do not present them here.
The resulting transmission and reflection amplitudes are computed from
\bea
r_n=\sum_m\vec C^{\,T}_n\check M_{nm}\vec B_m t_m,\label{r_n}\\
\sum_m \vec A^{\,T}_n \check M_{nm}\vec B_m t_m=\delta_{n0},\label{t_n}
\eea
where
\bea
\vec A_n &=& \left[ \frac{\bar\eta_n^L - \eta_n}      {\eta_n - \bar\eta_n} \spin(\bar\eta_n,-1) e^{-i\kappa_n L_1}
                           + \frac{\bar\eta_n - \bar\eta_n^L}{\eta_n - \bar\eta_n} \spin(\eta_n,-1)       e^{i\kappa_n L_1} \right]
\frac{e^{i\kappa_n^LL_1}}{\sqrt{2v_n^L}} ,\\
\vec B_n &=& \left[ \frac{\eta_n^R - \bar\eta_n}{\eta_n-\bar\eta_n} \spin(1,\eta_n)       e^{-i\kappa_n L_2}
                           + \frac{\eta_n - \eta_n^R}       {\eta_n-\bar\eta_n} \spin(1,\bar\eta_n) e^{i\kappa_n L_2} \right]
\frac{e^{i\kappa_n^R L_2}}{\sqrt{2v_n^R}},\\
\vec C_n &=& \left[ \frac{\eta_n - \eta_n^L}       {\eta_n - \bar\eta_n} \spin(\bar\eta_n,-1) e^{-i\kappa_n L_1}
                            + \frac{\eta_n^L - \bar\eta_n}{\eta_n - \bar\eta_n} \spin(\eta_n,-1)        e^{i\kappa_n L_1} \right]
\frac{e^{-i\kappa_n^L L_1}}{\sqrt{2v_n^L}} .
\eea
This system of equations for $t_n(k_y,E)$ reduces for the static case (then only $t_0$ is relevant) to Eq.~(\ref{t_static}).
For the case of no contact doping of the leads, i.e. $U(x)=0$, these equations reduce to
Eq.~(B14) in Ref.~\onlinecite{Korniyenko:2016ct}.

\end{widetext}

\section{Boundary condition Bessel function expansion}\label{app:M}

In this section we show that the boundary condition at the oscillating delta barrier in Eq.~(\ref{expM}) can be
rewritten in terms of Bessel-functions of the first kind.
The matrix elements $\check M_{nm}$ in Eq.~(\ref{t_n}) for transmission
amplitudes, which determines the strength of sideband coupling, thereby decay with
increasing $|n-m|$ as $J_{|n-m|}(Z_1)$.

The tensor $\check M$ in Eq.~(\ref{expM}) that represents the boundary condition at the delta barrier can be written as,
\be
\check M=\exp\Big[iZ_0\sigma_x\otimes\check \Gamma_0\Big]\exp\Big[i\frac{Z_1}{2}\sigma_x\otimes\check \Gamma_1\Big].
\label{bc_separated}
\ee
We will rewrite it to highlight the sideband space distribution. We will start by expanding the ac part of it in a Taylor series,
\be
\check M_{AC}=\exp\Big[i\frac{Z_1}{2}\sigma_x\otimes\check\Gamma_1\Big]=\sum\limits_{l=0}^{\infty}\left(i\frac{Z_1}{2}\sigma_x\right)^l\otimes\frac{\check \Gamma_1^l}{l!}.
\ee
Let us study the off-diagonal matrix $\check\Gamma_1$ taken to the $l$'th power, i.e. $\check\Gamma_1^l$.
Its matrix elements are given by binomial coefficients
\be
\left(\check \Gamma_1^l\right)_{nm}=\frac{l!}{\frac{l+|n-m|}{2}!\frac{l-|n-m|}{2}!}(l+1+|n-m|\mod2),
\label{binomial}
\ee
where $|n-m|\leq l$. Matrix elements for $|n-m|>l$ are zero.
Let us now introduce a matrix with unity entries on its $(\pm d)$'th diagonals,
\be
(\check\Gamma_d)_{nm}=\delta_{|n-m|,d}.
 \label{Gamma_diagonals}
 \ee
Note that $\check\Gamma_0$ and $\check\Gamma_1$ in Eq.~(\ref{bc_separated}) are included in this definition.
Then we can rewrite Eq.~(\ref{binomial}) by setting $d=|n-m|$. We obtain
\be
\check \Gamma_1^l=\sum\limits_{d=0}^{l}\frac{l!}{\frac{l+d}{2}!\frac{l-d}{2}!}(l+1+d\mod2)\check\Gamma_d.
\ee
The Taylor series is therefore given by
\be
\check M_{AC}=\sum\limits_{l=0}^{\infty}\sum\limits_{d=0}^{l}\frac{(i\frac{Z_1}{2}\sigma_x)^l\otimes\check\Gamma_d}{\frac{l+d}{2}!\frac{l-d}{2}!}(l+1+d\mod2)
\ee
By introducing a substitution $\tilde l=\frac{l-d}{2}$ we can rewrite it in a more convenient form
\be
\check M_{AC}=\sum\limits_{\tilde l=0}^{\infty}\sum\limits_{d=0}^{\infty}\frac{(i\frac{Z_1}{2}\sigma_x)^{2\tilde l+d}\otimes\check\Gamma_d}{(\tilde l+d)!\,\tilde l!}.
\ee
Using the Bessel function of the first kind series representation,
\be
J_d(Z_1)=\sum\limits_{l=0}^{\infty}(-1)^l\frac{\left(\frac{Z_1}{2}\right)^{2l+d}}{(l+d)!\,l!},
\ee
we arrive at
\be
\check M_{AC}=\sum\limits_{d=0}^{\infty}i^d J_d(Z_1)\sigma_x^d\otimes\check\Gamma_d.\label{eq:M_ac_bessel}
\ee
Including the dc prefactor, we arrive at
\be
\check{M}_{nm}=\exp[iZ_0\sigma_x](i\sigma_x)^{|n-m|}J_{|n-m|}(Z_1).\label{eq:M_bessel}
\ee

\end{appendix}


%

\end{document}